\documentstyle[12pt,lnfprep,epsfig]{article}
\begin{document}









\begin{titlepage}
\vskip 4cm
\vskip 2cm
{\large \bf Galactic Basins of Helium and Iron around the Knee Energy}
\vskip 0.4cm

\begin{center}
Antonio Codino$^1$, Fran\c{c}ois Plouin$^2$\\
\vskip 0.2cm
{\small \it ${}^{1)}$ INFN 
         and Dipartimento di Fisica dell'Universita degli Studi
         di Perugia, Italy.}\\
{\small \it ${}^{2)}$ former CNRS researcher, LLR,
         Ecole Polytechnique, F-91128 Palaiseau,France.}  
\end{center}

\vskip 1cm
\baselineskip=14pt
\begin{abstract}
The differential energy spectrum of cosmic rays exhibits
a change of slope, called $knee$
of the spectrum, around the nominal energy of $3 \times  10^{15}$ $eV$, and
 individual $knees$ for single ions, at different energies. 
The present work reports a detailed account of the characteristics and 
the origin of the knees for Helium and Iron.
Current observational data
regarding the magnetic field, the insterstellar matter density, the size of 
the Galaxy and the galactic wind, 
are incorporated
in appropriate algorithms which allow to simulate millions of
cosmic-ion trajectories in the disk. Bundles of ion trajectories  
define galactic regions called basins utilized in the present analysis of the knees. 
The fundamental role of the nuclear cross sections
in the origin of the helium and iron knees is demonstrated and highlighted. 
\par The results of the calculation are compared with
the experimental data in the energy interval  $ 10^{12}$ $eV$  - $ 5 \times 10^{17}$ $eV$.
There is a fair  agreement between the computed 
and measured energy spectra 
of Helium and Iron; rather surprisingly  their relative intensities 
 are also in accord with those  computed here. The results
suggest that acceleration mechanisms in the disk are extraneous to
  the origin of the $knees$.
\end{abstract}

\vspace*{\stretch{2}}
\begin{flushleft}
\vskip 0.2cm
{ This text appeared as the LNF report INFN/TC-06/05. See the link:}
\vskip 0.01cm
{ http://www.lnf.infn.it/sis/preprint/pdf/INFN-TC-06-5.pdf}
\end{flushleft}

\end{titlepage}

\pagestyle{plain}
\setcounter{page}2
\baselineskip=17pt


\section{Introduction}

In a recent companion paper  [1] the notion of galactic basin was adopted in order
to investigate some properties of galactic cosmic rays.
 This concept, introduced in a previous work [2], is extensively utilized in 
this study. The concept of galactic basin is similar to that of a terrestrial basin, 
peculiar of a river, with all the  $caveats$ inherent to any analogy.
When the position of an instrument recording the passage 
of cosmic ions
 and the source distribution in the disk are assigned, the galactic basin
is simply that particular region populated by the majority of the sources
feeding the instrument.
 Changing the position of the instrument,
the characteristics of the $basin$ (volume, extension, shape, contour, etc)  are
altered. 
 In our opinion, the concept of galactic basin is simple and straightforward,
and particularly useful when an efficient tool is required to compare,
investigate or even discover some properties of cosmic ions.

\par The purpose of this study is to identify those physical phenomena at the origin
of the $knee$ and to give a quantitative account of the major characteristics of the 
helium and iron $knees$. Their explanation is a part of a research effort
 to account for
 the knees and ankles of individual ions. The origin of the 
$knees$ presented here
are conclusively corroborated
in a   
forthcoming, collateral paper [3] demonstrating the interconnection
between the $knees$ and $ankles$ of  individual ions of the cosmic-ray spectrum.

\par The results and the prerequisites of this investigation are organized as follows.
In the second Section the hypotheses and the parameters 
involved in the calculation are enumerated. Since they have been presented
in previous studies [4,5,6],
 their description is concise. Only for two upgrades of the simulation code ($Corsa$),  
 the description
is ampler; these upgrades are introduced here for the first time. They regard the galactic wind and the 
propagation of the particles at energies greater than $ 5 \times 10^{11}$ $eV/u$,
the previous limit in the maximum energy of $Corsa$. In Section 3 the
 properties of 
galactic basins of Helium and Iron, in the interval
 $ 10^{12}$ and $ 5 \times 10^{17}$ $eV/u$, are summarized and discussed. 
Through many examples, it is investigated how the shape of basins change
 with increasing energy.
In Section 4 are clarified and consolidated 
some results given in Section 2.
In Section 5 it is investigated in detail
how nuclear cross sections affect the change of the spectral index of individual ions.
The interaction cross sections
of Helium and Iron with the insterstellar matter 
are set artificially constant. Then the differential energy
spectrum of Helium and Iron is studied by counting the number of
cosmic rays intercepting the local galactic zone.
In Section 6
the influence of the magnetic field and the galactic wind
on the results previously obtained is calculated and
discussed. 
In Section 7 a comparison of the predicted and measured 
 energy spectra  of Helium and Iron  is presented. 
Unfortunately, 
even  a concise comparison (to preserve a tolerable length of the paper) 
 with the experimental data necessitates
logical and theoretical prerequisites,
 partially presented in Section 7, partially
described elsewhere [7,8].

\par Protons are not included in this analysis, in spite of their importance, 
because, at very high energy,  computational procedures become quite complex and 
very time consuming. 
Primary protons interacting with the interstellar hydrogen
yield many secondaries, indistinguishable from primaries, which make the
analysis of the results quite involved  as reported in a previous study limited
to the energy of $ 10^{11}$ $eV$ [4]. The present calculation deals only with
 Helium and Iron because the behavior of all other ions lighter than Iron,
 except protons, fall between 
these two nuclei, as previously demonstrated [1].
The energy interval considered extends from  $ 10^{12}$ to $ 5 \times 10^{17}$ $eV/u$,
 which is adequate for the $knee$ problem.

\section{Observational data incorporated in the simulation algorithms}

\par
\begin{figure}
\centerline {\epsfig {figure=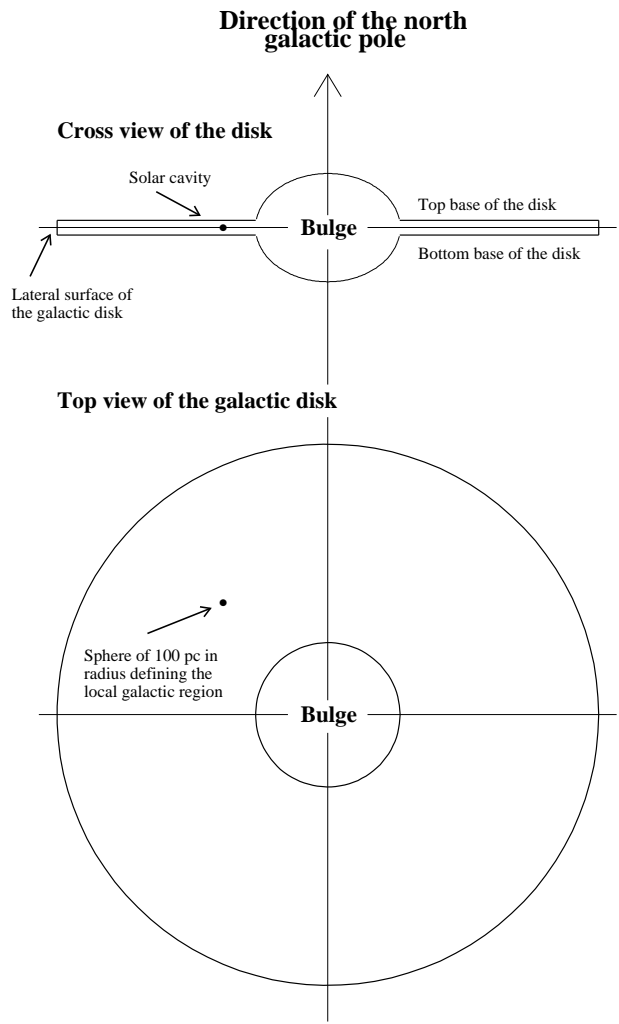,width=12cm}}
\caption{\it Lateral and top view of the galactic disk indicating
the local zone and the Bulge.
\label{fig:fig1}}
\end{figure}

\begin{figure}
\centerline {\epsfig {figure=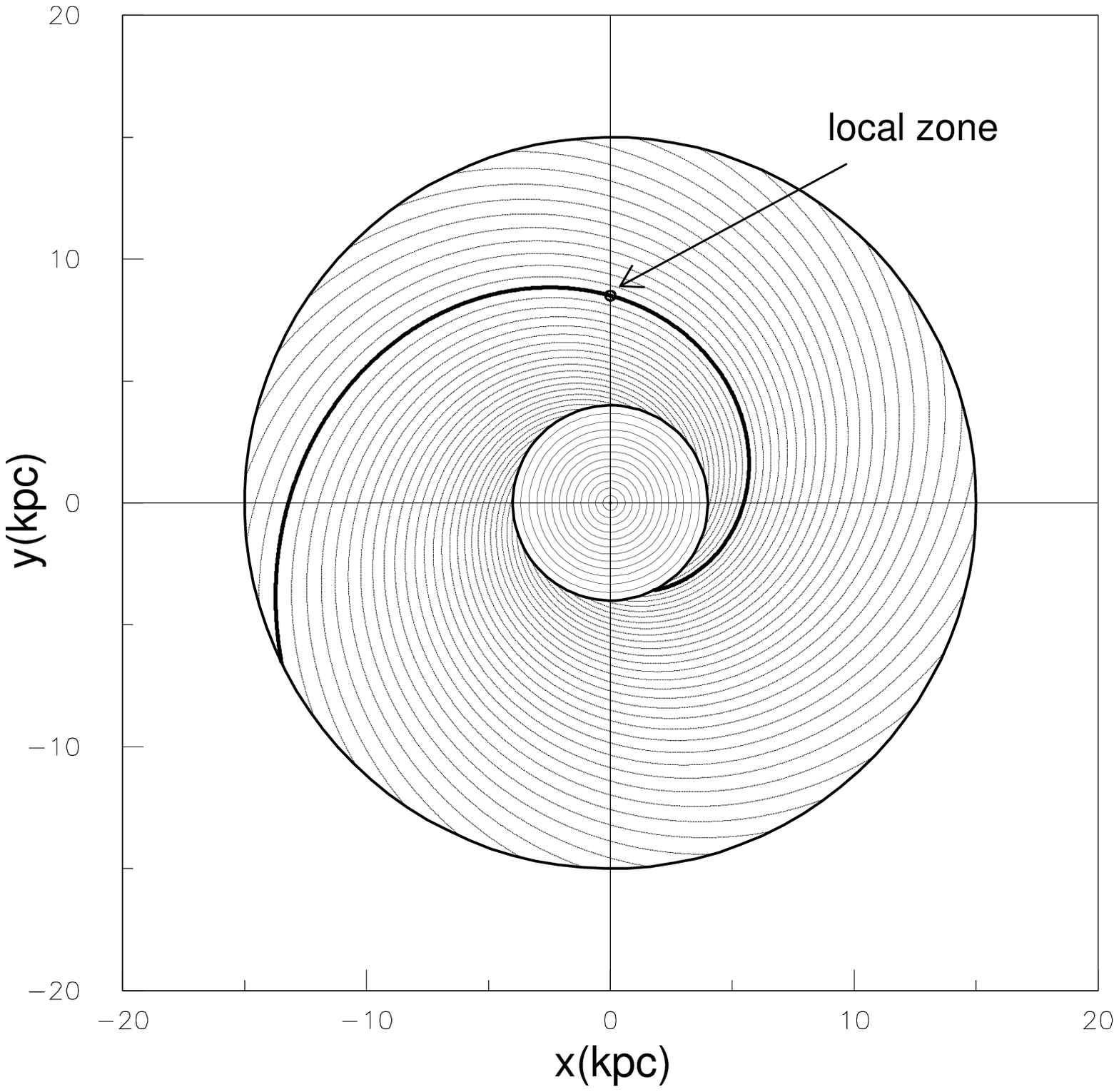,width=9cm}}
\caption{\it Pattern of field lines of the spiral magnetic field projected
onto the galactic midplane.
The spiral field line (thick line) departing from the bulge,
intercepting the local galactic zone
and terminating on the disk frontier ($r$ = 15 $kpc$) is called 
the principal field line.
\label{fig:fig2}}
\end{figure}

\begin{figure}
\centerline {\epsfig {figure=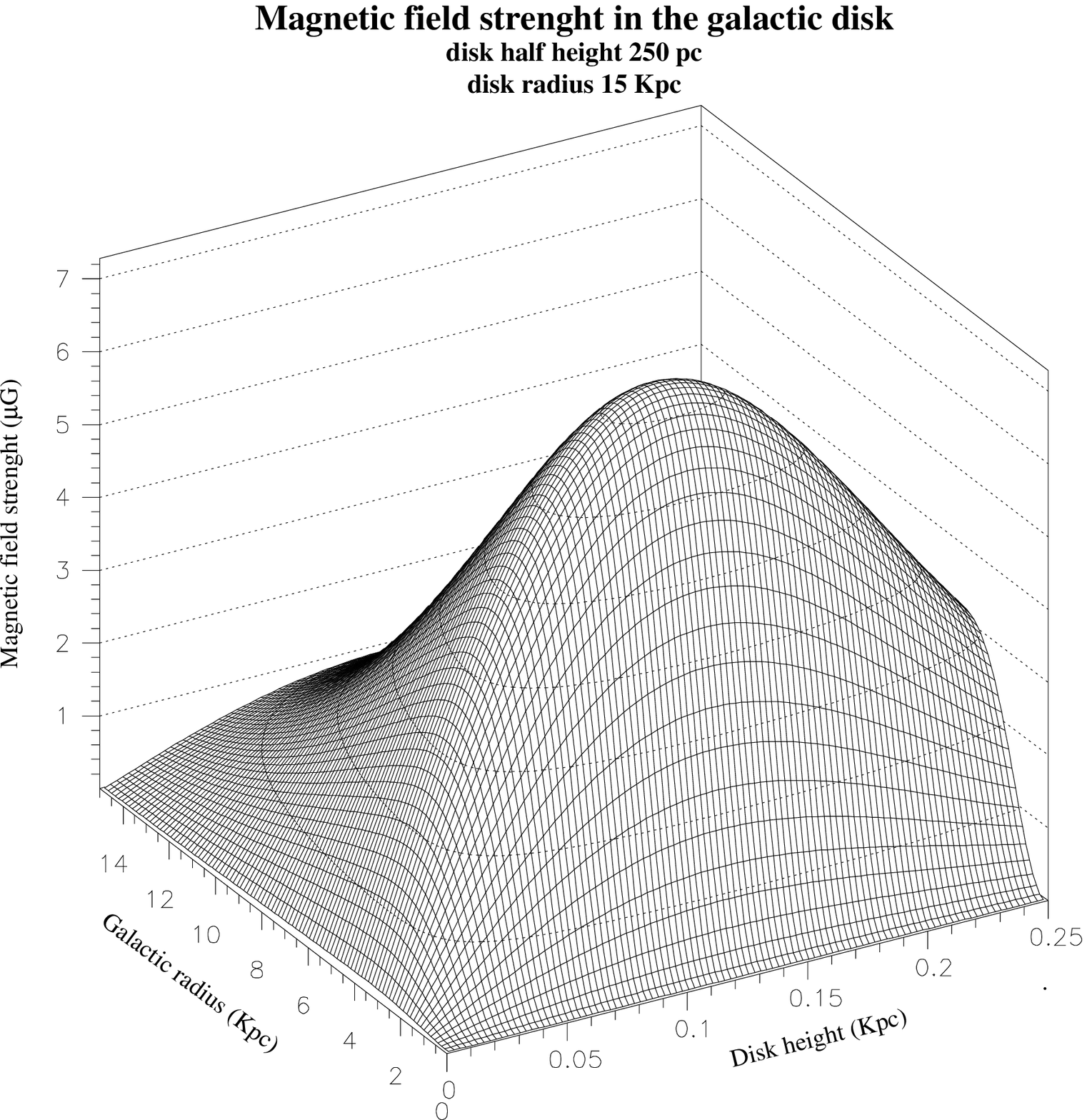,width=9cm}}
\caption{\it Magnetic field strength in the disk. 
 Note that the axis of the disk height has a different scale from that 
representing the disk radius.
\label{fig:fig3}}
\end{figure}

 In the following the set of parameters utilized in the calculation are 
concisely presented. The form and dimensions of the Galaxy are displayed in
fig.$\,$\ref{fig:fig1}.
A cylindrical frame of reference is used with coordinates $r$, $z$ and $\phi$.
The Bulge is an ellipsoid with a major axis  of $8$ $kpc$, lying onto the
galactic midplane,
and a transverse minor axis of 6 $kpc$. The disk has a radius of $15$ $kpc$ and a half 
thickness of $250$ $pc$. 
The interstellar matter consists of pure hydrogen with constant density
of 1 atom $per$ $cm^3$ [9] enhanced to 1.24 to take into account heavier nuclei. 
The galactic magnetic field is a superposition of regular
 and  chaotic fields. In fig.$\,$\ref{fig:fig3}
are shown the line pattern of the spiral field
and the principal field line (thick line) which intercepts the Earth
at coordinates $r$=8.5 $kpc$, $z$=$0$ $kpc$ and $\phi$=$90$ $degrees$.
The field strength of the regular component 
is shown in fig.$\,$\ref{fig:fig2}. 
The coherence length of the regular field is $125$ $pc$ [12].   
The chaotic field is materialized
by magnetic cloudlets [ see, for example, references 10,11] of spherical form
 with a mean radius of 2.5 $kpc$ fluctuated
by a uniform distribution with a width of $0.5$ $pc$. The ion 
propagation normal to the regular spiral field
is generated by the chaotic field leading to a transverse-to-longitudinal displacement
ratio of 0.031,
in a range compatible 
with the quasi-linear theory of ion propagation
in an astrophysical environment [13].

\par The source distribution,  $Q(r,z,\phi)$, regarded as uniform, is the following :
$$ Q(r,z,\phi) = C \theta(r-R) N(\sigma,z) \quad \quad \quad (1)$$
where $C$ is a normalization constant, $\theta$ is the radial distribution
with a disk radius $R$ = $15$ $kpc$ and  N($\sigma$,$z$) is the vertical distribution
with a gaussian shape with standard deviation $\sigma$ = $80$ $pc$.
The function  N($\sigma$,$z$)  makes
smooth the source distribution along $z$
with respect to a steep, unphysical, $z$-rectangular 
distribution. The function   $Q(r,z,\phi)$ is not dissimilar
 to the positions of the galactic sources distributed as
supernovae remnants [14,15] with the advantage of
avoiding a specific, constraining hypothesis on source positions.
Replacing  N($\sigma$,$z$) by  $\theta(z-z_0)$ in equation (1), 
a strictly uniform distribution is obtained where $z_0$ is the disk half thickness.
The mean distance of the source positions from the local zone is denoted by 
$D_{si}$.

\par The flattening
of the electron spectrum in external galaxies,
detected by radiotelescopes at
frequencies about 1 GHz, is commonly regarded as observational evidence for
the existence of the galactic winds [16]. This evidence is less
convincing in the Milky Way because, the synchrotron radiation detection and analysis 
of the results are quite complex, being the Earth embedded in the Galaxy. 
A three dimensional formula, based on 
observational data,  for the
galactic wind in the disk and halo volume, is not known.
 Usually, a one dimensional
 wind along $z$ with a linear velocity 
is considered in many investigastions of 
cosmic rays, though a radial wind component
 is not excluded [17]. From these inputs it seems appropriate for this calculation a linear 
formula of the type:
$$ v(z) = a z \quad \quad \quad \quad \quad \quad (2)$$
where $a$ is a suitable constant constrained by the maximum wind 
velocity, $v_{max}$, at the disk boundaries.
Simulation algorithms shift along $z$ the trajectory segments
by the amount $\delta_z$ according to the equation (2), 
namely, $\delta_z$ = $v$($z$)$\tau$ where $\tau$ is the time interval taken by 
 the cosmic ion to travel the length $l_z$ of the trajectory segment along $z$.
 The maximum wind velocities
adopted in a variety of cosmic-ray investigations in the disk, span the range
5 to 40 $km$/$s$.
 The account of the Boron-to-Carbon 
flux ratio at very low energy [18] 
probably represents an incisive example of the 
role of the galactic wind
on cosmic rays. 
Instead of adopting a particular value of $v_{max}$, which is unnecessary in this work
as it will be clear in Sections 6 and 7,  the parameter $a$ and the related maximum wind velocity,
 is left as a free parameter.

\par New algorithms have been added for the propagation of cosmic rays around
the knee energy. The maximum energy of the previous version of $Corsa$ 
was at $ 5 \times 10^{11}$ $eV/u$.
\begin{figure}
\centerline {\epsfig {figure=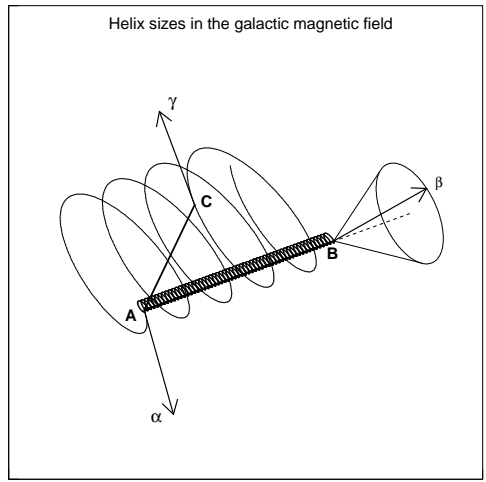,width=9cm}}
\caption{\it Qualitative illustration of the sizes of three helices in the
galactic magnetic field.  The field is aligned
along the straight line  $AB$ and it can be a chaotic or
 a regular field segment.
 A particle of low energy, initially at the point $A$ with velocity $\alpha$, 
will closely follow 
the magnetic field, with a small helix radius, reaching the point $B$
with a velocity $\beta$.
In this case the velocity vectors
$\alpha$ and $\beta$ are randomly oriented (fully uncorrelated). 
A particle of sufficient high
energy at $A$ with velocity $\alpha$, describing the ample 
helix from $A$ to $C$, leaves the point $C$  
with the velocity vector $\gamma$. In this case, unlike the previous one, 
the vectors $\gamma$ and $\alpha$ are fully correlated.
 The vast majority of cosmic rays, having very low energy compared
to the $knee$ energy, simply follows the regular field lines with a
small tranverse displacement and their helix radii are so small to be 
indistinguishable from the segment $AB$ ( thick line ). 
\label{fig:fig4}}
\end{figure}
 Fig.$\,$\ref{fig:fig4} reports, as an example, 
three helical trajectories winding round the magnetic field, whose direction
coincides with the segment $AB$, $150$ $pc$ long.
The radii of the three helices are 0.01, 4 and 50 $pc$ in  a uniform field 
with $3$ $\mu$$G$ strength. The smallest helix is indistinguishable with
the line segment $AB$, indicative of the fundamental, standard  condition of
 cosmic ions in the Galaxy at energies below $10^{14}$ $eV/u$;
while travelling in the Galaxy they are mainly bound to the lines of the 
regular field. Details of the propagation procedure at very high
energy are given in the Appendix $B$.

\par Particles which enter the Halo are regarded as lost, since the albedo
is less than 2 per cent at the energy of $ 10^{11}$ $eV$[1] and decreases 
with the energy. Either a  circular or a spiral 
field can be used in the Halo with appropriate strength (see equation (3) of ref.[1]). 
Matter 
density in the Halo is  a factor $10^2$-$10^3$ less than that in the disk. 
In the present investigation, however, the influence of the Halo and its size,
 is quite marginal.

\section{Galactic basins around the knee energies}

\par In this Section the major characteristics of the galactic basins 
are summarized. Note that the total length of the principal magnetic 
field line from the Bulge ($r$ = 4 $kpc$ ) to the disk radius ($r$ = 15 $kpc$) is 
43.5 $kpc$.
\par
\begin{figure}
\centerline {\epsfig {figure=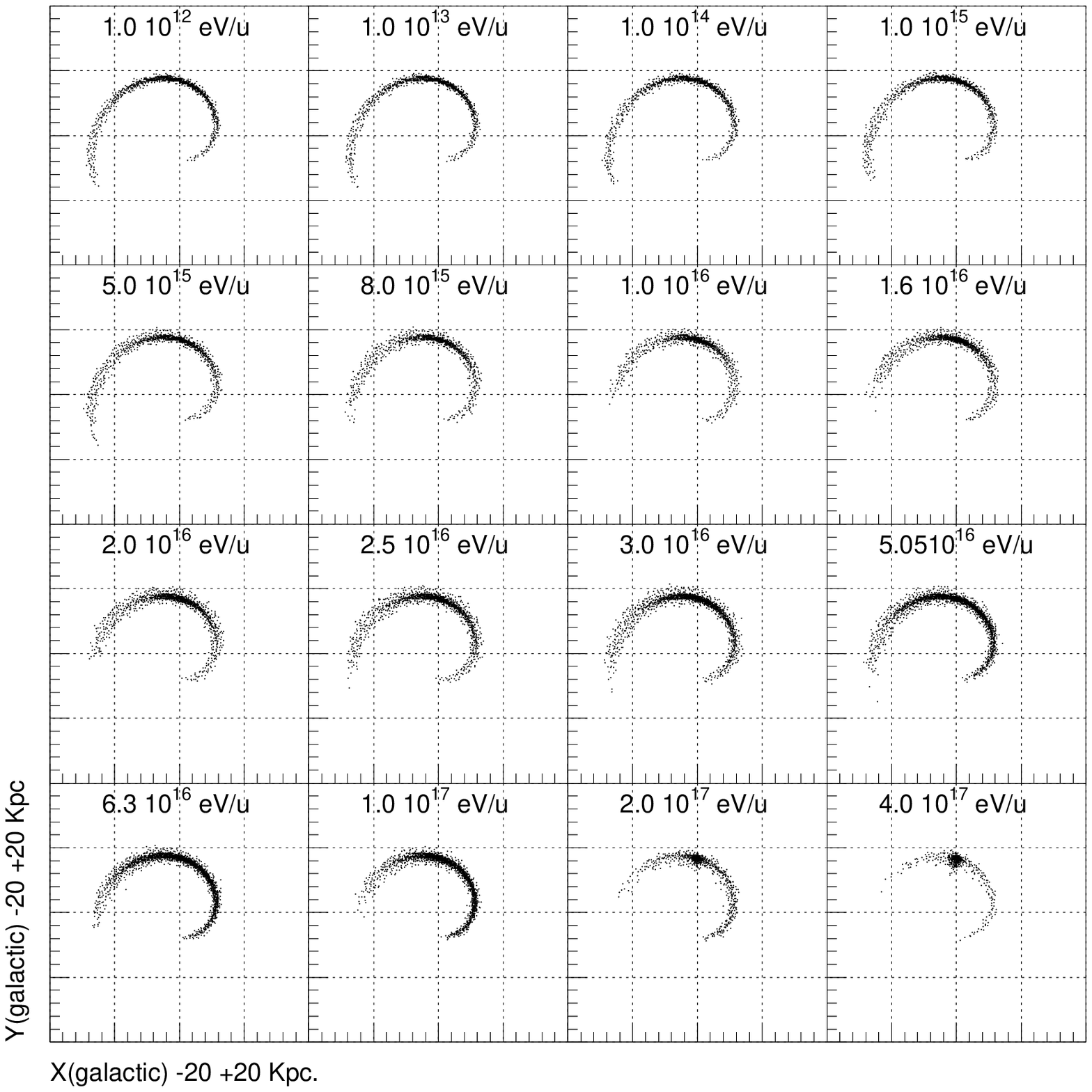,width=15cm}}
\caption{\it  Different forms of the galactic basins projected onto
the galactic midplane   
illustrating   how they
 modify as the energy increases from $ 10^{12}$ to $ 4 \times 10^{17}$ 
$eV$/$u$. Each figure derives from an initial
sample of $2 \times 10^5$ ions. The frame of reference and the position
of the local zone are those shown in figure 2.
\label{fig:fig5}}
\end{figure}

 In fig.$\,$\ref{fig:fig5} are shown the positions of the Helium sources
having trajectories intercepting the local galactic zone at sixteen
different energies (from  $ 10^{12}$ to
$ 4 \times 10^{17}$ $eV/u$). The source positions are projected onto the
galactic midplane. The sequence of the plots highlights the changes of the 
source distributions
with increasing energy.
 The corresponding spatial distributions of the iron sources (not shown)  have 
similar structure, though the values of the parameters characterizing the basins
are, of course,  different.
 Source positions populate a narrow strip, along 
the principal magnetic field line, except at the maximum energy of 
$4 \times 10^{17}$ $eV/u$ 
where the basin form is radically different from the previous ones.
 This fundamental feature was 
identified at low energy below $ 10^{11}$ $eV/u$, long time ago by ourselves, for proton
and beryllium ions,
and recently for all other ions from Helium to Iron [1].
 These characteristics of
the ion basins derive, ultimately, from  
observational data
of the Milky Way, and particularly, from the spiral shape of the 
magnetic field shown in fig.$\,$\ref{fig:fig2}. As a general rule, spiral galaxies possess
a regular magnetic field with a spiral shape, Andromeda being a 
notable exception with an annular field. This feature, together with
the small ratio of transverse-to-longitudinal ion propagation, yields
long and narrow basins.

\par
\begin{figure}
\centerline {\epsfig {figure=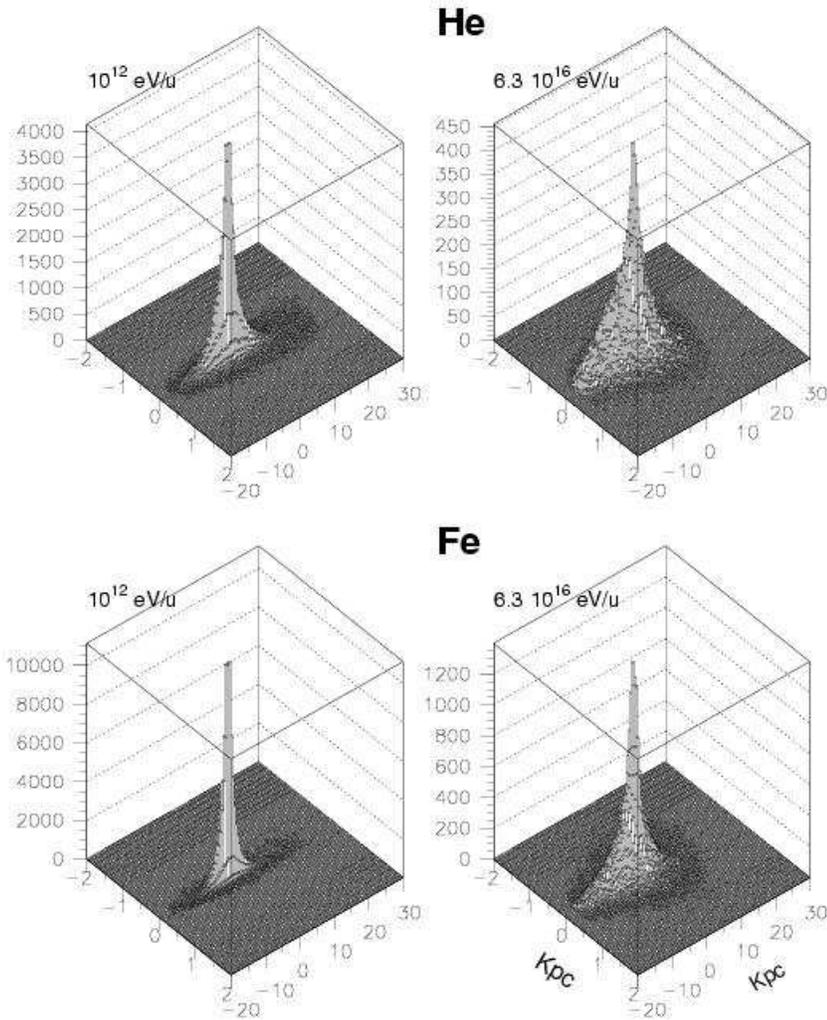,width=12cm}}
\caption{\it Spatial distribution of helium and iron sources in the Disk
at the energies of  $10^{12}$ and $ 6.3 \times 10^{16}$ $eV/u$. The results are
 obtained with an initial
sample of $200$ $000$ trajectories for both ions. The source arrays of the 
highest energy are typical of  the $knee$ energy region.
\label{fig:fig6}}
\end{figure}

  Fig.$\,$\ref{fig:fig6} 
displays a tridimensional view onto the galactic midplane 
of the positions of helium and iron sources emanating trajectories
which intercept the local galactic zone 
in a reference frame 
where the horizontal axis is the principal field line.
 This coordinate system and its 
advantages are described in  Appendix A. Each plot contains samples of $ 2 \times 10^5$
ion trajectories. To illustrate the basin profiles two arbitrary energies 
 of $10^{12}$ and  $ 6.3 \times 10^{16}$ $eV/u$ are chosen. 
At the energy of  $10^{12}$ $eV/u$ both basins are well
contained in the Disk, the iron basin being more concentrated
around the local zone
because its trajectory lengths are shorter than those of Helium.
 At the energy of  $10^{16}$ $eV/u$
both basins  increasingly populate a zone toward the bulge frontier. Note that 
in fig.$\,$\ref{fig:fig6} the bulge lies at coordinate $-16.7$ $kpc$ along the principal field 
line, (i.e. the axis of the figure) while the disk edge at $24.1$ $kpc$.

\par
\begin{figure}
\centerline {\epsfig {figure=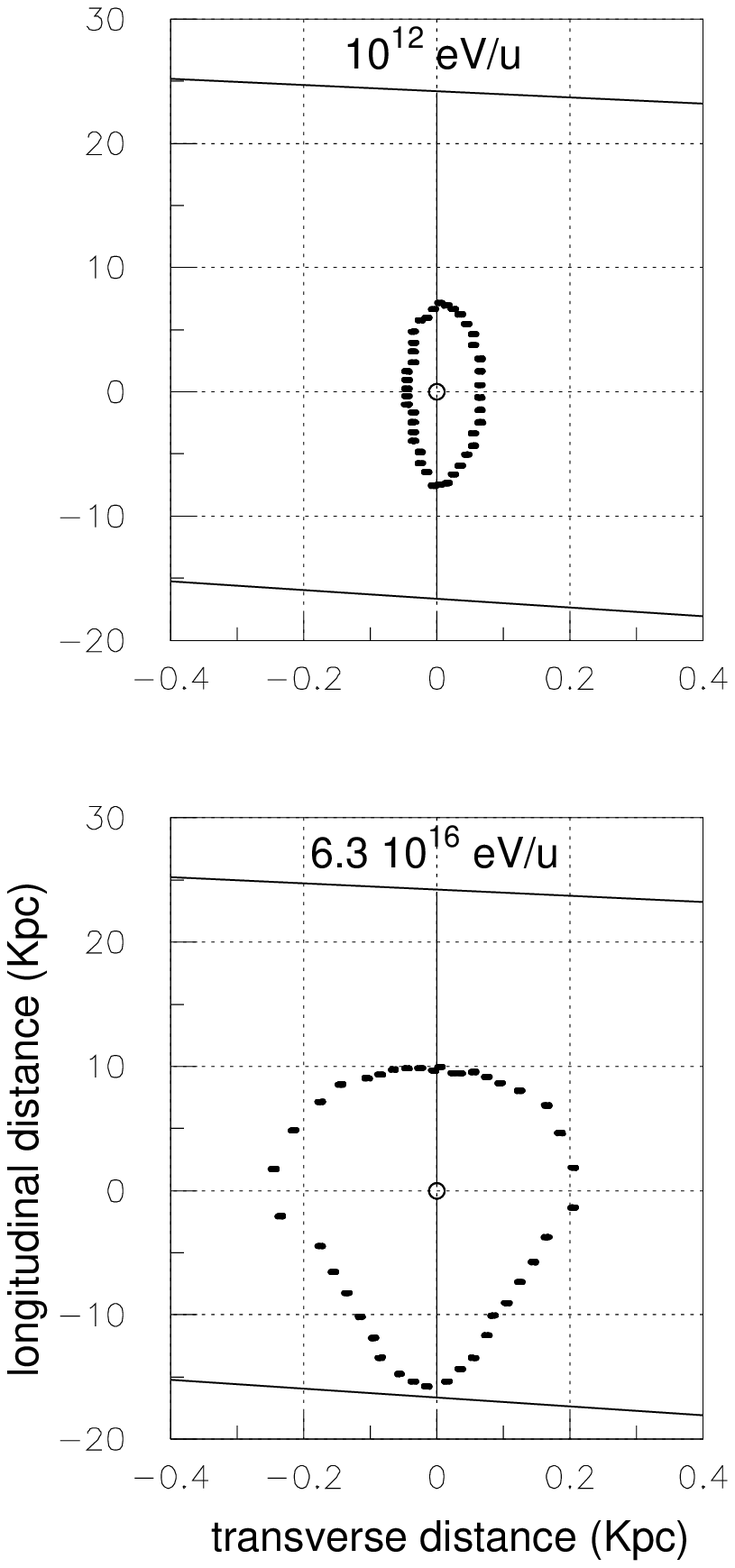,width=11cm}}
\caption{\it Contour plots containing $90 \%$ of iron sources
at the energy of  $10^{12}$ and $ 6.3 \times 10^{16}$ $eV$. The contours
refer to same iron samples  shown in  fig.$\,$\ref{fig:fig6}.
The energy of the lower contour  lies in the $knee$ energy region.
\label{fig:fig7}}
\end{figure}

 The spatial distribution of the sources 
in fig.$\,$\ref{fig:fig6} can be visualized by contour plots
containing $90 \%$ of the sources.
The arbitrary value of this percentage  was previously adopted (see figure 
5 of reference [1] ).
 Fig.$\,$\ref{fig:fig7} shows the contours of two Iron basins at the
energies of  $ 10^{12}$ and 
$ 6.3 \times 10^{16}$ $eV/u$.
The lower and upper solid lines in fig.$\,$\ref{fig:fig7}  
delimit, respectively, the Bulge and the disk internal frontiers.
In order to better emphasize the transversal structure
of the basin, the horizontal axis in fig.$\,$\ref{fig:fig7} 
is magnified by a factor 50.
 
It is evident that the basin shape changes from an ovoidal form,
typical at low energy,  to a ram head, typical at high energy.
 These alterations of the basin shape suggest that  cosmic-ray populations
suffer difforming filtering and processing as the energy varies. 
 It is apparent, for example, the basin enlargement
with increasing energy. 
 The supposed continuity between the two contour
plots in fig.$\,$\ref{fig:fig7}, 
passing from a narrow, ovoidal shape to a large ram-head shape, 
manifests the transverse enlargement with the increasing energy.
Note also 
that the basin at 
$ 10^{12}$ $eV/u$ is well contained in the Disk, with longitudinal ends approximately equidistant
from the bulge and disk frontiers. 
 On the contrary, at the energy of $ 6.3 \times 10^{16}$ $eV/u$ the longitudinal 
ends of the basin
are not equidistant from the bulge and disk frontiers.
 The relative high clustering of cosmic-ion 
sources  closer to the Bulge frontier ($r$ = 4 $kpc$) is due to  the magnetic field strength, 
quite large in this zone ($4 - 6 \mu$$G$) compared to the 
strength ($0.5 - 1 \mu$$G$) around
the disk edge ($r$ = 15 $kpc$) as indicates fig.$\,$\ref{fig:fig3}.

\par The basin extension   is characterized by the length $L_B$, along the 
principal field line,
 and  the transverse width $W_B$. The changes in
 the form of a basin are quantified in terms of  $L_B$ and $W_B$.
\begin{figure}
\centerline {\epsfig {figure=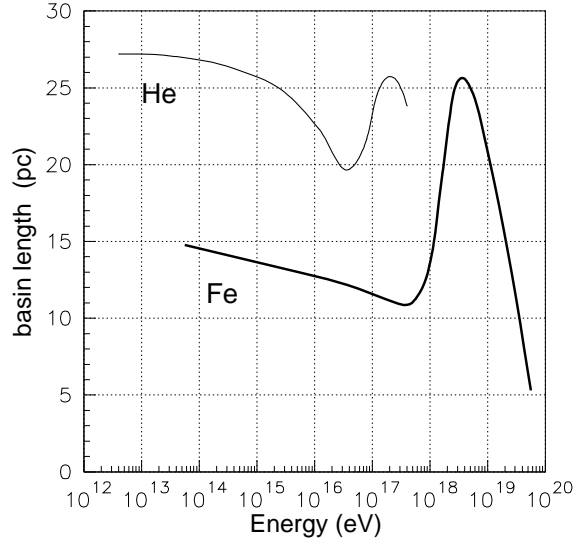,width=8cm}}
\caption{\it Basin length 
as a function of energy for Helium and Iron. The helium minimum of $18$ $kpc$ 
occurring
at the energy of $ 4 \times 10^{16}$ $eV$ is related to the $knee$ position
along the energy axis. Similarly it takes place for Iron, with a minimum of 10.5 $kpc$ 
at the energy
of $ 6 \times 10^{17}$ $eV$ 
\label{fig:fig8}}
\end{figure}
 Fig.$\,$\ref{fig:fig8} reports 
$L_B$ $versus$ energy for Helium and Iron. Note that the extensive air shower experiments 
 measure the total kinetic energy of the ions, therefore it is useful,
here and in all subsequent plots, to adopt this unit of measurement. For both 
ions $L_B$ gradually decreases
up to a minimum occurring, approximately,  at the energy of $ 4 \times 10^{16}$
 $eV$ for Helium and at $ 5 \times 10^{17}$ $eV$ for Iron.
After the minimum, $L_B$ increases to a maximum value, 26 $kpc$ for Helium and
25 $kpc$ for Iron, then it decreases again. The $knees$ of the two ions are
related to the minima of $L_B$ as it will be shown later. While the decrease of $L_B$ 
for Iron is perfectly linear in the 
interval $ 10^{13}$-$ 10^{17}$ $eV/u$,
that of Helium is slightly arc shaped; this feature is accounted for
by the larger extension of helium basins compared to those of Iron, which lose
helium ions from the large basin periphery.
The general decrease of $L_B$ with energy is due to the rising cross sections.
The analysis of the structure of the maxima of $L_B$ 
in fig.$\,$\ref{fig:fig8} is beyond the perimeter of this paper and it will be
discussed in a forthcoming paper.

 \par For the sake of completeness let us mention that the ratio $L_B$/ $W_B$ has 
been also calculated as a function of the atomic weight
and shows a regularity,  useful in some circumstances (see figure 5 of ref. [1]).

\section{Illuminating the galactic Disk by an ion beam}

Illuminating the Galaxy with an imaginary isotropic beam emitted in the local
zone and counting the number of nuclear collisions occurring in the Disk, 
the characterization of the galactic basins around the $knee$ energies is
greatly facilitated. In fact, the number of trajectories simulated by this
procedure is about two orders of magnitude higher than the direct method used in Sections
5, 6 and 7. The direct method utilizes the source positions  $Q(r,z,\phi)$, 
and counts the number of particles reaching the local zone, $n_g$.

\par The simulation algorithms reconstruct cosmic-ion trajectories
originating in the Disk. They are classified  according to the physical process 
causing the ion to disappear from the disk:
nuclear collisons, ionization energy losses,
and escape trajectories. By escape is meant trajectories rooted 
in the disk but terminating into the Halo.
A generic sample of cosmic rays will contain the fractions of cosmic trajectories
denoted respectively $f_n$, $f_i$ and  $f_e$  with the obvious condition:  $f_n$ + $f_i$ 
+ $f_e$ = 1. This partitions were adopted in all 
previous works made using the code $Corsa$. 
Above $ 10^{11}$ $eV$, the ionization energy loss is thoroughly negligible, hence $f_i$= $0$
and a simplified, useful situation appears, where    $f_n$ 
+ $f_e$ = 1, and therefore the knowledge of $f_n$ entails that of $f_e$.

\par Typical samples of $ 2 \times 10^5$ trajectories for each energy
have been simulated 
and therefore the quantity $f_n$ is simply the number of nuclear
collisions in the disk divided by $ 2 \times 10^5$.  
\begin{figure}
\centerline {\epsfig {figure=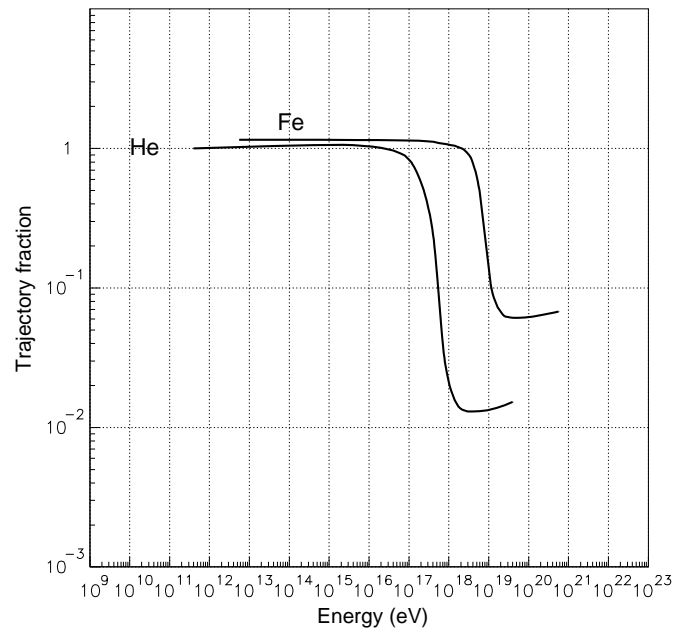,width=9cm}}
\caption{\it Fraction of helium trajectories terminated
in the disk by nuclear collisions as a function
of energy from a sample of $ 2 \times 10^5$ trajectories at each 
energy (for a set of 25 energy values). 
A distinctive feature of the curve is a high plateau, a fall and a minimum
with a subsequent tiny rise (low plateau). The gap between the high 
and the low plateau
is due to the effect of nuclear cross sections.  
\label{fig:fig9}}
\end{figure}
 Fig.$\,$\ref{fig:fig9} gives
$f_n$  as a function of the energy for Helium and Iron.
At the energy of $ 10^{12}$ $eV$ the fraction of nuclear collisions
in the disk for Helium is set arbitrarily to 1. Iron collisions are normalized accordingly. 
Typically, for $ 2 \times 10^5$ ion sources in the disk 
with an energy of $10^{14}$ $eV$ there are  $ 1.828 \times 10^{5}$ 
 collisions for Helium and  $ 1.992 \times 10^{5}$  for Iron; 
 therefore in this case, $f_n$= $1.055$ for Helium and  $f_n$= $1.150$ for Iron, if
the number of helium collisions at $ 10^{12}$ $eV$  is  $ 1.63221 \times 10^{5}$.

 \par The helium fraction $f_n$ has a tiny rise in a large energy interval, 
$ 10^{10}$-$ 10^{16}$ $eV$, reaching a
maximum of $1.058$ at $ 1.6 \times 10^{15}$ $eV$ ( hardly visible in the figure), 
 then it falls steeply 
in the energy decade $ 10^{17}$-$ 10^{18}$ $eV$. The fall is correlated with
 the helium knee as it will be apparent later.
 The helium fraction $f_n$ increases by $5.8\%$ in the 
range $ 10^{12}$-$ 10^{16}$ $eV$.
 Since  $\sigma$(He) and $\sigma$(Fe) increase with energy so does the number of
collisions in the disk.
 This is obvious from the expression of the inelastic
collision length, $\lambda$:
$$ \lambda =  { A \over { \sigma  d  N_a} }   \quad \quad \quad (3)$$
where A is a suitable atomic weight of the interstellar medium, $d$ its density and
$N_a$ is the Avogadro number.
 In fact, for a constant disk thickness, 
a uniform matter density
and a constant number of sources,
a lower $\lambda$ entails a higher number of collisions.

\par The decrease of  $f_n$ 
by $50\%$ with respect
to its maximum value in the interval  $ 10^{10}$-$ 10^{16}$ $eV$ identifies
(rather arbitrarily but conveniently ) the particular energies where ion losses 
from the disk become dominant, since $f_e$ = $1$ - $f_n$. These values are 
$ 2.2 \times 10^{17}$ $eV$ for Helium and $ 5.5 \times 10^{18}$ $eV$ for 
Iron.

\par The influence of nuclear cross sections and of the magnetic field 
strength on $f_n$ can be separated by setting, artificially, $\sigma$(He)
 and $\sigma$(Fe), at constant values.
\begin{figure}
\centerline {\epsfig {figure=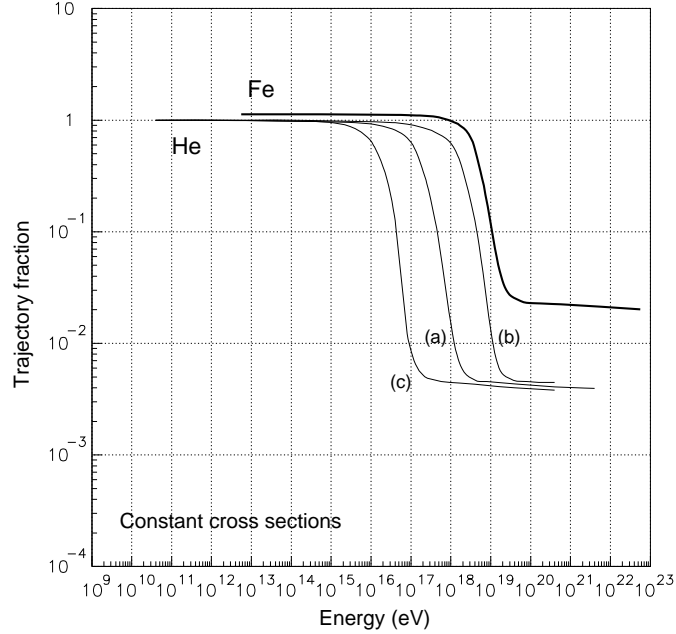,width=9cm}}
\caption{\it Fraction of cosmic-ray trajectories terminated
in the disk by nuclear interactions. 
For these results constant cross sections $\sigma$ are used. 
The striking effect of the magnetic field on $f_n$ is evident in the 
three curves $a$, $b$ and $c$. Unlike the results shown in figure 9, here $f_n$
is strictly constant, before the fall due to constant $\sigma$. 
\label{fig:fig10}}
\end{figure}
 The results are shown in fig.$\,$\ref{fig:fig10} 
which gives $f_n$ $versus$ energy, using constant cross sections, namely,
$\sigma$(He)= $115$ $mb$ and $\sigma$(Fe)= $775$ $mb$, in the full energy range
(see also Section 5). 
In this case, unlike the results given in fig.$\,$\ref{fig:fig9},
  $f_n$ is rigorously constant in the range
 $ 10^{11}$-$ 10^{15}$ $eV$, as expected.

\par The abrupt decrease of $f_n$ with the energy manifests the
inefficiency of the galactic magnetic field to retain
cosmic helium, with a uniform efficiency, beyond $ 10^{17}$ $eV$. Adopting a 
fictitious magnetic field with
a field strength 10 times higher than that shown in fig.$\,$\ref{fig:fig2},
the thin curve $a$ transforms into the curve $b$.  The curve $c$ refers to a magnetic field 
strength attenuated by a factor 10. The beautiful separation 
between the curves $a$, $b$ and $c$, and the 
ensuing 
shifts in the positions of the bends, along the energy axis, mark the role of the 
magnetic field in the origin of the $knees$.
Note that $f_n$ attains a minimum value,
rather constant beyond $ 10^{18}$ $eV$, with a characteristic gap between the
high  and the low plateau, amounting to $ 4.5 \times 10^{-3}$   for Helium and 
 $ 2.2 \times 10^{-2}$  for Iron,
as shown in fig.$\,$\ref{fig:fig10}.
 The height of the gap is a peculiarity of the Galaxy, the ion and the nuclear cross sections.
The two minima signal the onset of the rectilinear propagation.

\par Note also the interesting phenomenon that at low energy, below 
$ 2 \times 10^{15}$ $eV$ and at very high energy, above $ 10^{18}$ $eV$,  the
three curves $a$, $b$ and $c$ coincide.
 At very high energy the rectilinear propagation
of cosmic ions sets on and the bending power of the magnetic field cease to be efficient, 
thoroughly vanishing above $ 10^{19}$ $eV$.
 At low energy below $2 \times 10^{15}$ $eV$, ions experience multiple bendings 
and multiple inversions of the motion, regardless
of the particular energies of the ions,
 and the three curves coincide again.

\par The corresponding curves for Iron show a similar behaviour, and particularly, the 
triple splitting of $f_n$ with the three values of  magnetic field strength
is  observed.

\par For the sake of completeness and subsequent convenience
\begin{figure}
\centerline {\epsfig {figure=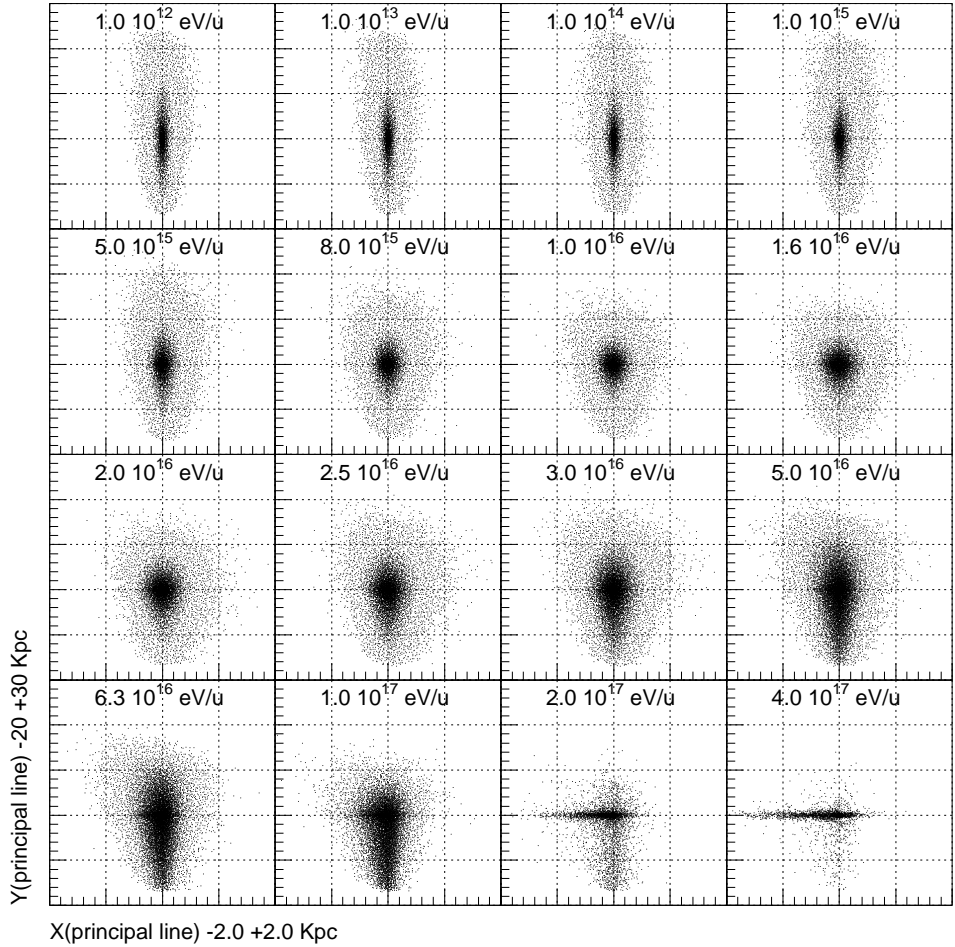,width=15cm}}
\caption{\it Spatial distributions of helium sources at sixteen
different energies in a frame of reference  (described in the Appendix $A$)
where the vertical axis coincides with the principal magnetic field line.
 Note the characteristic deformations as the energy increases and the radical changes
in the basin forms around the knee energies $ 10^{16}$ $eV$/$u$ and those
 marking the onset of the rectilinear propagation above  $ 10^{17}$ $eV$/$u$ .
\label{fig:fig11}}
\end{figure}
in fig.$\,$\ref{fig:fig11} are given the source distributions for Helium along the 
principal field 
line intercepting the Earth (see fig.$\,$\ref{fig:fig2}), at sixteen different energies. 
The transverse structure of the source positions is more evident 
in this frame of reference than in that used in fig.$\,$\ref{fig:fig5},
 which better emphasizes the
role of the regular field to channel cosmic ions. It is 
particularly evident the 
change of the source positions around the $knee$ energy band, where basins
become larger and larger (see fig.$\,$\ref{fig:fig8}) 
and devoid of sources 
(the falls in fig.$\,$\ref{fig:fig9} and fig.$\,$\ref{fig:fig10}).
 The minimum of the basin length,
apparent in fig.$\,$\ref{fig:fig8}, occurs at the  energies 
between $ 10^{16}$ and $ 2 \times 10^{16}$ $eV$. It is impressive the alterations 
in the source distributions around the energy of $ 10^{16}$ $eV$/$u$ which signal
the presence of the helium knee. 
 The source positions in fig.$\,$\ref{fig:fig11}
complements those given in fig.$\,$\ref{fig:fig5}.

\par The results obtained in fig.$\,$\ref{fig:fig11}, interpreted as source positions
and not $per$ $se$,  
take advantage of the
trajectory reversibility in a magnetic field. An approximate method of calculation
discussed and checked previously [1] which allows to simulate a large number of trajectories. 
Typical samples of $2 \times 10^5$ cosmic rays
are injected from the local galactic zone and the corresponding trajectories 
reconstructed through the disk volume. The coordinates of the nuclear collisions have
 been recorded. In this specific calculation the initial point of the 
trajectory is the local zone, while the final point coincides with the position
 of the inelastic nuclear collision. By inverting the initial and final point 
of the trajectory the source distribution feeding the local zone is calculated. 
This calculation procedure implicitly preassumes a source distribution strictly uniform 
in the disk and not that described by equation (1).

\section{The influence of the nuclear cross sections}

\par The magnetic field and the nuclear cross sections, $\sigma$, 
determine the positions of the fall of $n_g$ along the energy axis i.e. the positions
of the individual $knees$.  In order to single out the effects of these
two parameters, the nuclear cross sections have been artificially set at a constant
value (115 mb for Helium and 775 mb for Iron), above the energy 
of $ 5 \times 10^{11}$ $eV$.
With this artifice, the influence of
$\sigma$ on the knee becomes more patent. As it is well known proton-proton and 
nucleus-proton   cross sections increase with energy beyond $ 4 \times 10^{11}$ $eV$,
where a vague minimum occurs,  up to the
maximum energy 
reached in accelerator experiments, $ 8 \times 10^{16}$ $eV$ [19].
\begin{figure}
\centerline {\epsfig {figure=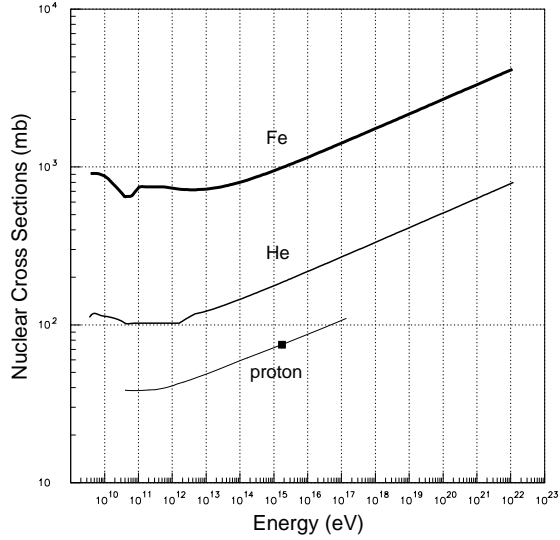,width=9cm}}
\caption{\it Inelastic cross sections Helium-proton and
Iron-proton $versus$ energy used in the calculation. The proton-proton
cross sections is also shown with a data point at the maximum energy
obtained in accelerator experiments.
The cross section of 38.8 $mb$
at about $40$ $GeV$ is an average value of many experiments.
\label{fig:fig12}}
\end{figure}
 Fig.$\,$\ref{fig:fig12} shows the rise of nuclear cross 
sections with energy for Helium $\sigma$(He), Iron $\sigma$(Fe) and proton
$\sigma$(p). As an 
example, at $40 GeV$/$u$  $\sigma$(He)
 is $103$ $mb$ and $\sigma$(Fe) = $720$ $mb$, 
 while at  $ 10^{17}$ $eV$  $\sigma$(He)
 is $306$ $mb$ and  $\sigma$(Fe) = $2054$ $mb$.

\par
\begin{figure}
\centerline {\epsfig {figure=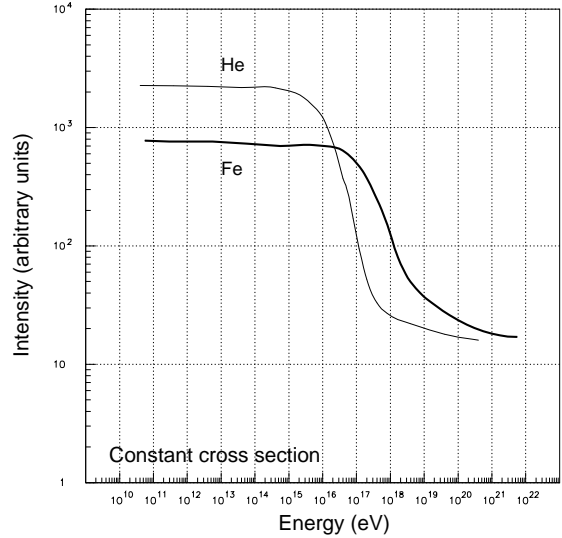,width=9cm}}
\caption{\it Number of cosmic ions intercepting the local galactic zone, $n_g$,
versus energy for Helium and Iron using constant cross sections in the
entire energy range.
\label{fig:fig13}}
\end{figure}

 Using the spatial distribution of the sources  $Q(r,z,\phi)$ 
and reconstructing the trajectories of cosmic ions,
the number of particles intercepting the local galactic zone, $n_g$ is counted. 
The quantity $n_g$ is directly proportional to the differential energy 
spectrum of the cosmic rays, $dn$/$dE$. 
 Fig.$\,$\ref{fig:fig13} gives $n_g$ for Helium and Iron as a function of energy.
 Since equal samples 
( tens of millions) of Helium and Iron trajectories have been 
simulated for the results given in fig.$\,$\ref{fig:fig13},
the powers of the sources for the two ions in the galactic disk are equal. 
In Section 7, in the
comparison with the experimental data, this irrealistic hypothesis, though 
convenient here, is removed.
 The results in fig.$\,$\ref{fig:fig13}
indicate that the  
iron intensity is a factor 2.93 lower than that of Helium in the energy band
where $n_g$ for both ions is approximately constant. It turns out that $n_g$ for Helium 
is constant up to the energy of $2 \times 10^{15}$ $eV$, 
then a rapid fall initiates. In the energy band $ 8 \times 10^{15}$-$5
\times 10^{17}$ $eV$ the fall of $n_g$   
exhibits a constant slope with an index of 3.27. The same 
trend of $n_g$ 
is displayed
for Iron with a spectral index of 3.08, thus the curve 
is unevenly shifted in energy with respect to the
 helium curve.

\par The approximate constancy of $n_g$ for Helium in the interval 
$ 10^{12}$ - $10^{15}$ $eV$
suggests that the particle transport  
across the local zone attains somehow a stationary regime.
 Note that the disk
boundaries are sufficiently
remote from the local zone so that $n_g$ remains almost unaffected.
 Neither play a role
effects related to the simulation algorithms such as the diameter of the local zone
or the representation of the helices (see Appendix B). 
The particular energy of the bend
in fig.$\,$\ref{fig:fig13} directly manifests the effect of the magnetic field 
on cosmic ions reaching the local zone, depurated from the influence of rising cross sections.

\par The energies where the bends occur, $ 6 \times 10^{15}$  $eV$ for Helium
and $ 2 \times 10^{17}$ $eV$  for Iron, remind of the results reported
in fig.$\,$\ref{fig:fig8} 
regarding the lengths of the longitudinal axes of the basins, $L_B$.
 A notable circumstance
is that the energies characterizing the bends in fig.$\,$\ref{fig:fig13}, 
are the same where
the minima of $L_B$ in fig.$\,$\ref{fig:fig8}  occur.
This is not surprising because the volume of the galactic basins, and in turn $L_B$, 
is certainly interrelated to  the particle  intensities $n_g$.
 Taken this circumstance into account,
the same considerations developed in Sections 4 and 5 apply on the interplay
of the cosmic-ion 
populations around the knee.

\par For an assigned energy of the ion and $\sigma$,
the positions of the bends are intimately related 
$in$ $primis$ to the form and mean strength of the magnetic field and $in$ $secundis$
to the thickness of the disk. By altering one of these two parameters
the bend positions  critically change.
 Thus, as an example, setting a 
fictitious magnetic field,
10 times higher than the real field, 
changes similar to those shown in fig.$\,$\ref{fig:fig10} -- are calculated.
 Using a completely different simulation code 
a comparable effect is observed [8].

\par
\begin{figure}
\centerline {\epsfig {figure=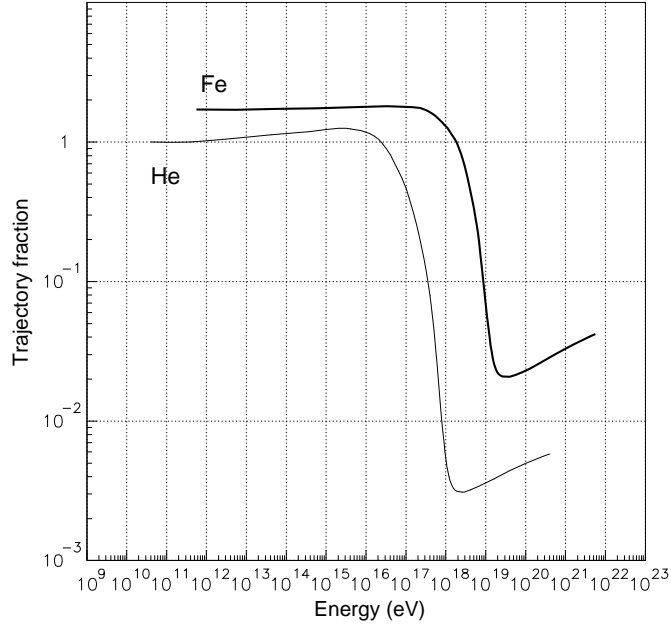,width=9cm}}
\caption{\it Fraction $f_n$ of helium trajectories terminated
in the disk by  nuclear collisions as a function
of energy assuming the spatial distribution of sources $Q(r,z,\phi)$
and rising cross sections. The normalization of $f_n$ is the same adopted
in figure 9.
\label{fig:fig14}}
\end{figure}

 It is also interesting to evaluate the number of nuclear collisions 
occurring in the disk, $f_n$,  with the source distribution $Q(r,z,\phi)$, 
 instead of a pointlike source
(the local zone) as performed in Section 4. 
 Fig.$\,$\ref{fig:fig14} reports $f_n$ versus energy, in this condition,
 for rising cross sections.

\par The helium curve reaches a maximum of 1.24 at  $ 4 \times 10^{15}$  
 $eV$ (the iron maximum is 1.80 at  $ 5.6 \times 10^{16}$ $eV$ ) which differs
 from the corresponding maximum of 1.058 at  $ 1.6 \times 10^{15}$ in figure 9.
The different amounts of interstellar matter encountered by cosmic ions 
for the two source positions, pointlike and  $Q(r,z,\phi)$,
account for the difference and, in some respects, quantify the influence
 of the matter thickness on the energy spectra.

\par From this analysis comes out that
the fall of $n_g$ with the energy, beyond   $ 6 \times 10^{16}$ $eV$  for Helium
and above $10^{17}$ $eV$  for Iron, is due to a decreasing efficiency
of the Galaxy to retain particles as the energy increases.

\par Two competing mechanisms  govern the trend of $n_g$ $versus$ energy.
 Let us analyze, for example, how they operate in the knee region. As displayed in 
Fig.$\,$\ref{fig:fig8}, helium basins
enlarge above the energy of  $ 4.0 \times 10^{16}$ $eV$ where the minimum of
$L_B$ occurs. Consider two energies
$E_1$ and $E_2$ with $E_2 > E_1$ and the corresponding basins denoted by $B_1$ and  
$B_2$, respectively (the corresponding volumes denoted also  $B_1$ and $B_2$). 
Assume further  that $E_2$ is in an energy range located in the fall of $n_g$.
Since the basin volumes  enlarge with energy   $B_2  > B_1$. Let us
further denote by $D_B$ 
the disk volume obtained by removing the volume of the basin $B_1$
from that of $B_2$ ($D$ in $D_B$ is for difference in the basin volumes). 
Let $n_1$ be the number of sources in $B_1$ at energy $E_1$ and $n_2$  that in
 $B_2$  at energy $E_2$. Then, subdivide $n_2$ into two contributions 
such that $n_2$ =  $\delta_1$ + $\delta_2$, 
where $\delta_1$ is the
number of sources of the basin $B_1$ at the energy $E_2$ and $\delta_2$ is
the number of sources in the volume $D_B$  at the energy $E_2$. \par Of critical importance
for the origin of the $knees$ is the quantity ($n_1$ - $\delta_1$) in comparison 
with $\delta_2$. The basin enlargement from $E_1$ to $E_2$ 
allows 
new sources, located in $D_B$,  to increase the value of $n_g$ by 
the amount $\delta_2$ (first mechanism).  
When the energy reaches $E_2$ some sources in $B_1$
do not emanate cosmic-ray trajectories intercepting the local zone, 
and consequently $n_g$ tends to decrease by the amount ($n_1$ - $\delta_1$) since 
   $n_1 > \delta_1$    (second mechanisn).
This happens because, as the energy increases, trajectories are less twisted
by the magnetic field and therefore they have minor
probability of reaching the local zone. The decreasing intensity $n_g$,
in fig.$\,$\ref{fig:fig13},
 in the energy range
 $10^{15}$ - $2.0 \times 10^{17}$ $eV$, 
indicates that the second mechanism overwhelms the first one, i.e. ($n_1$ - $\delta_1$)
$\gg$ $\delta_2$. 
The approximate constancy 
of $n_g$ in the interval
 $ 10^{11}$ - $10^{15}$ $eV$ indicates that the two mechanisms approximately balance
i.e. ($n_1$ - $\delta_1$)
$\sim$  $\delta_2$. 
Note
that in this case $B_1 > B_2$.

\par The interplay of the two mechanisms
becomes more intricate when real cross sections (i.e. rising cross sections) are inserted.
In this case, the interplay of cosmic-ion populations with energy
persists, as described above, but the mean length of ion trajectories decreases with energy,
due to the rising cross sections as it is obvious from equation (3).

\section{The effect of the rising cross sections, the galactic wind and the magnetic field}

\par Before facing a comparison with experimental data it is useful to further explore
some characteristic features of the basins around the knee energy. Basins undergo
radical changes 
in the knee energy region and therefore those variables characterizing the basins, 
$L$, $g$, $L_B$, $D_{si}$, $n_g$ and  $f_n$ should absorb somehow the upheaval.

\par
\begin{figure}
\centerline {\epsfig {figure=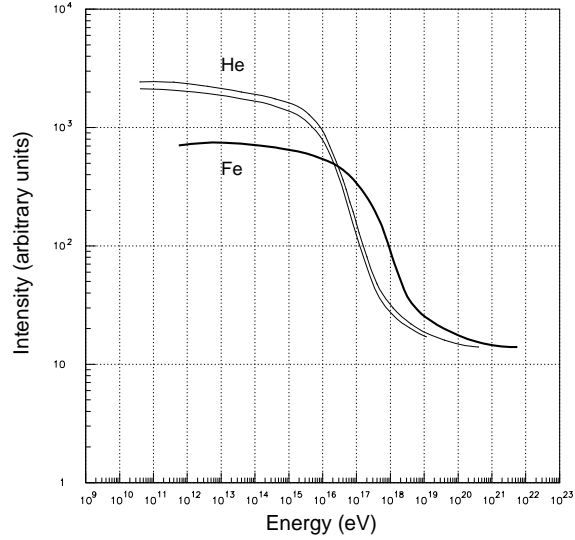,width=9cm}}
\caption{\it Number of cosmic ions intercepting the local galactic zone, $n_g$,
versus energy for Helium (solid line)  and Iron (thick line) with rising  cross 
sections.
 The normalization of $n_g$ is the same adopted in fig.$\,$\ref{fig:fig13}.
 The effect
of the galactic wind on the helium spectrum is represented by the thin line.
The wind effect on the iron spectrum is negligible and it is not shown.
\label{fig:fig15}}
\end{figure}

 In the following, constant cross sections are replaced by the rising cross 
sections, shown in fig.$\,$\ref{fig:fig12}. 
 Fig.$\,$\ref{fig:fig15} shows $n_g$ $versus$ energy for Helium and Iron.
 These results are
of great importance because they are directly comparable with the experimental
data. Comparing the corresponding curves of $n_g$
in fig.$\,$\ref{fig:fig13} (constant
cross sections), the positions of the bends 
in fig.$\,$\ref{fig:fig15} are placed  at lower energy. 
In addition, $n_g$ $versus$ energy around the bend is smoother
than the corresponding curve in fig.$\,$\ref{fig:fig13}.
 There are energy regions around the maximum fall of intensity
where the spectral indexes are constant amounting to $3.43$ for Helium 
and $3.38$ for Iron.

\par With exactly the same trajectories used in the calculation of 
fig.$\,$\ref{fig:fig15},
 the helium and iron $grammage$, $g$, 
as well as the mean distance of the sources from the local zone, denoted $D_{si}$,
 are determined. 
\begin{figure}
\centerline {\epsfig {figure= 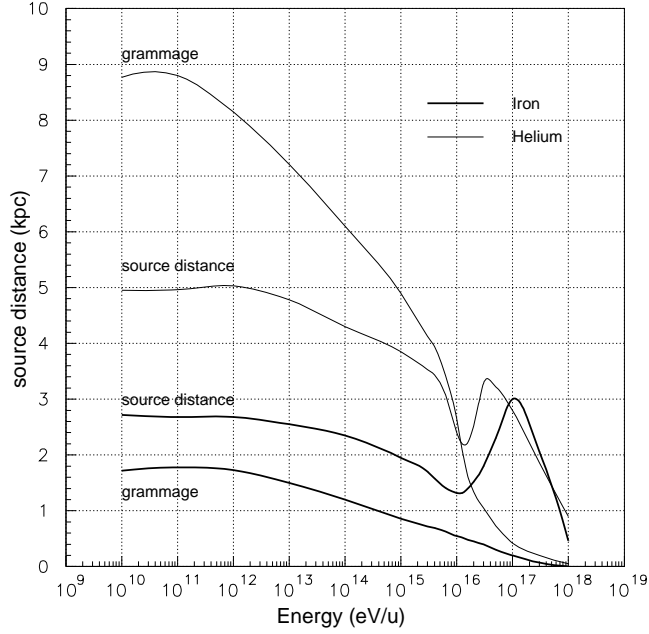,width=9cm}}
\caption{\it Iron and helium grammage versus energy in the same conditions
of fig.$\,$\ref{fig:fig15}. The mean distance
of the sources from the Earth, $D_{si}$, is also given. This latter quantity
presents two characteristic peaks accounted for by the radical alterations
in the cosmic-ion populations around the knee energy region. It is
a fortuitous occurrence that the vertical axis expresses the grammage
in units $g$/$cm^2$ and the source distance in $kpc$ as well. 
\label{fig:fig16}}
\end{figure}
These two quantities are shown in fig.$\,$\ref{fig:fig16} 
and they complement previous results
on the geometrical characterization of the basin. 
The grammage is related
to the trajectory length, $L$,  by the equation : $g$ = $m$ $n$ $L$ where $n$ is the
number of atoms per $cm^3$ in the interstellar space and $m$ the mass
of the average atom in the insterstellar medium.
The helium grammage decreases almost parabolically as the 
energy increases, up to the energy of $ 10^{16}$ $eV$, attaining an asymptotic limit
above $ 10^{18}$ $eV$ for both ions.

\par The quantity $D_{si}$ versus energy shows a broad peak with a maximum 
at the energy of $3 \times 10^{16}$ $eV$ for Helium and around $ 10^{17}$ $eV$ for Iron. 
 It may seem both surprising and inexplicable that  $D_{si}$ has a peak while $g$ none. 
 Comparing the results of 
fig.$\,$\ref{fig:fig15} and fig.$\,$\ref{fig:fig16} 
 it is evident that the peak energies lie
where the maximum steepness  
of $n_g$ occurs. These peaks are explained in the 
light of the results
described in Sections 3 and 4  and by direct analysis of the simulated 
trajectories.
 In fact, the mean 
distance of the sources from the local zone diminishes with the energy,
as the tails of the basins gradually disappear, as apparent from
fig.$\,$\ref{fig:fig5}  and fig.$\,$\ref{fig:fig11}. 
The progressive loss of the tails
corresponds to a transverse enlargement of the basin which, in turn, 
implies a reduced distance from the sources.
 Fig.$\,$\ref{fig:fig8}  shows that the length of the basin, $L_B$,  
increases
beyond the minimum at the energy of $ 4 \times 10^{16}$ $eV$, then it reaches a maximum
to finally  decrease again. 
As the basin length increases beyond the minimum, some new distant sources add to 
the enlarging basin, enhancing  $D_{si}$, 
which tranlates in a bump visible in fig.$\,$\ref{fig:fig16}.

\par The grammage is dominated by long trajectories rooted close to the local
 zone but slightly aside the principal field line.
Trajectories along the principal field line have much small lengths.
Long trajectories are the result of multiple inversions of motion
in the magnetic field, and not of long distances between the sources and the 
local zone. Since the energy region of the two peaks in  $D_{si}$
corresponds to the onset of the rectilinear propagation, the expected
enhancement of $g$ or $L$ (expected only on the basis of that of $D_{si}$ )
 does not take place.

\par The effect of the galactic wind on the quantities $g$ and  $n_g$ 
 has been investigated,  in a plausible interval
of the wind velocity, and it turns out to be quite small but relevant to the aim
of this calculation. 
The smallness is due to the particular
position of the local zone e.g. $z$=$0$ $pc$ assuming, of course, galactic sources.  
In spite of its smallness, the inclusion in the algorithms significantly improves the agremeent
with the experimental data, as it will be shown in the next Section.
\begin{figure}
\centerline {\epsfig {figure=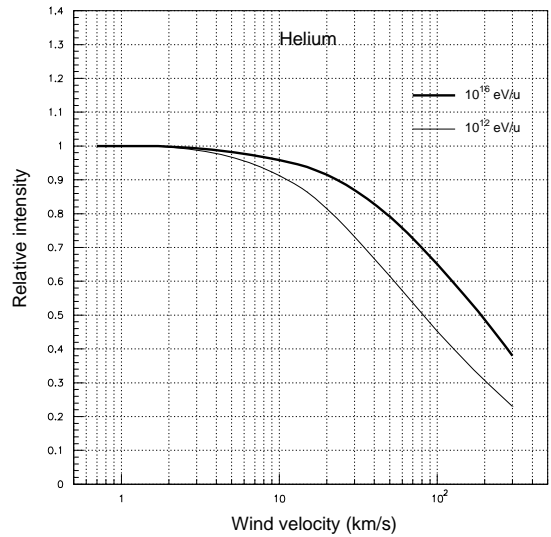,width=9cm}}
\caption{\it Number of helium trajectories reaching the local zone $n_g$ as a function
of the wind velocity at the energy of $10^{12}$  $eV/u$ (thin line) and $10^{16}$ 
 $eV/u$ (thick line).
The curves are based on samples of $ 2 \times10^{5}$  trajectories 
for each velocity value
for a total of $3$ millions trajectories.
 At zero wind velocity $n_g$ is set at $1$.
\label{fig:fig17}}
\end{figure}
 Fig.$\,$\ref{fig:fig17} reports $n_g$ for Helium versus wind velocity 
up to a maximum value of $300$ $Km$/$s$ (unphysically high in the disk ) 
for two arbitrary energies of $ 10^{12}$ and $ 10^{16}$ $eV$.
 Ions reaching the local zone decrease by about 
$10 \%$ with a wind velocity of $12$ $Km$/$s$.  
 The decrease is almost absent for Iron.
 This is due
to the mean length of iron trajectories, smaller than that of Helium, as it is evident
from equation (3) given the difference of $\sigma$(He) and $\sigma$(Fe).
\begin{figure}
\centerline {\epsfig {figure=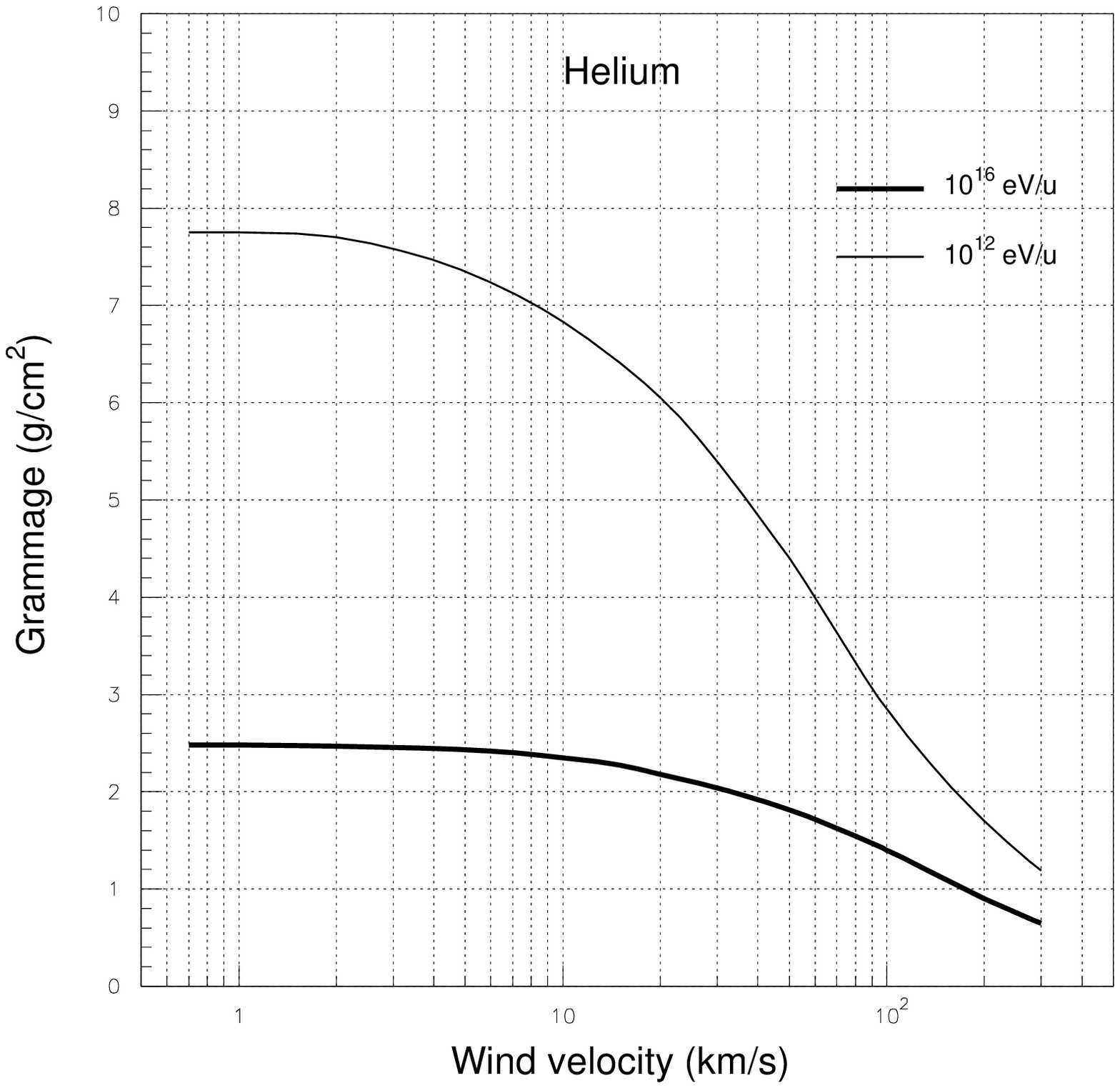,width=9cm}}
\caption{\it Grammage of helium trajectories reaching the local zone as a function
of the wind velocity at the energy of 
$10^{12}$  $eV$ (thin line) and $10^{16}$  $eV$ (thick line).
The curves are based on the same trajectory samples used 
for the results shown in fig.$\,$\ref{fig:fig17}.
\label{fig:fig18}}
\end{figure}
 Fig.$\,$\ref{fig:fig18}  gives the grammage versus wind velocity, for the same energies of the
previous figure.
 A tangible effect of the wind on the grammage starts above a
 wind velocity of $12$ $Km$/$s$. Accordingly comparable effects are expected
on the basin extensions;  they would be hardly visible in
figures like fig.$\,$\ref{fig:fig5}  and fig.$\,$\ref{fig:fig11}.

\par The parameters defining the influence of the magnetic field on relevant
properties of galactic cosmic rays  were the first to be 
tested in 1995, namely, the residence time of cosmic ions in the Galaxy as a functions 
of the diameter of the cloudlets [6]; later, the age and grammage of protons against the 
shape of the regular field (circular, spiral,
and elliptic) were calculated [4]. The Beryllium age against the shape of 
the regular field was also investigated [20]. The results of these calculations  
indicate that the magnetic field
structure and the other parameters of $Corsa$ conform, with the correct order
of magnitude, to relevant experimental data at low energy, especially
those regarding the age and grammage of cosmic ions.

\par
\begin{figure}
\centerline {\epsfig {figure=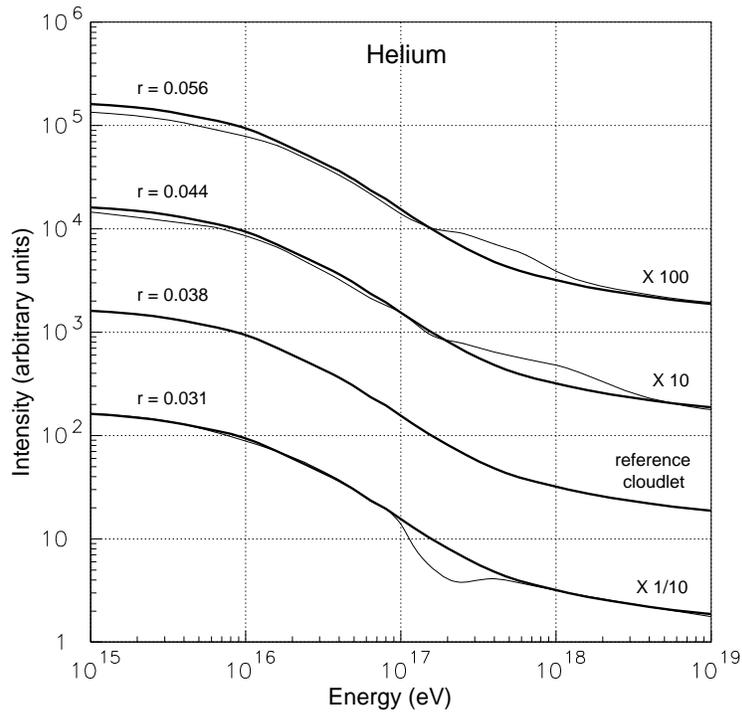,width=10cm}}
\caption{\it Helium trajectories crossing the local zone, $n_g$, versus energy
for different diameters of the magnetic cloudlets corresponding
a longitudinal-to-transverse displacement ratio of the cosmic ions 
$r$  from $0.031$ to $0.056$. 
The thick line, referred as standard cloudlet with
$r$ = $0.38$, is the same curve shown in fig.$\,$\ref{fig:fig15} 
stretched
in energy to better underline possible deformations.
\label{fig:fig19}}
\end{figure}

 The ratio of tranverse-to-longitudinal propagation, $r$,  in the magnetic and matter
structure of the Galaxy incorporated in $Corsa$ is close to the theoretical value
of $0.2$ - $0.4$ dictated by the quasi-linear theory of cosmic-ion propagation [13]. In fact,
the algorithms have a ratio of $0.031$ obtained by dividing the diameter of the magnetic 
cloudlets
to the coherence length of $125$ $pc$ and finally averaging over the field directions
(a factor $\pi$/$4$). The effect of this
fundamental parameter on 
$n_g$, e.g. the quantity directly related to the experimental data regarding 
the $knee$, has been analyzed.
 Fig.$\,$\ref{fig:fig19} reports $n_g$
for different cloudlet diameters, corresponding to alterations of 
$r$ from $0.031$ to $0.056$.
 Note that at very high energy the ratio $r$ should necessarily increase compared
to the ratio at low energy,
otherwise the ion gyroradius would be inferior to the mean transverse displacement
in the regular field, which is untenable. Thus, small values of $r$ are unphysical
at very high energy.

\par At small $r$ ions suffer a major effect
of the regular field, being easily channeled toward the disk periphery instead
of escaping along the $z$ direction,
creating a localized depression in the curve $n_g$ versus energy.
 This depression not only 
disappears for greater ratios $r$, as apparent from fig.$\,$\ref{fig:fig19}, 
but an opposite
phenomenon emerges, consisting in a small enhancement of $n_g$ above 
$ 6 \times 10^{17}$  $eV$. This is due to the
fragile conditions in the particle transport at these energies, 
close  to the rectilinear propagation, for
 which a major cloudlet size  
 translates into a higher ion flux
in the local zone. In conclusion, neither 
the bend positions in the energy axis nor the spectral index in the fall, are
significantly modified by the cloudlet radius around the reference value of 
$3$ $pc$ adopted in this calculation.

\section{Comparison with the experimental data}
It is  well known that the differential energy spectrum of the cosmic radiation
conforms to a power law like $a$$E^{-\gamma}$ where $a$ is a constant,
$E$ is the kinetic energy of the ions and $\gamma$ the spectral index. By the term
$knee$ is meant both the change in the spectral index from $2.7$ to $3.0$ and the nominal
energy where this change occurs, at about $ 3 \times 10^{15}$  $eV$. A third basic
peculiarity, not analyzed in this study, is the phase in the arrival 
direction of cosmic ions at Earth, which has been unambiguously
measured below and above the $knee$ energy (see, for example ref.[21]) displaying
a characteristic modification.

\par The $knee$ was 
discovered in 1958 [22]. This fundamental observation refers to all particles 
of the spectrum hereafter denoted $complete$ $spectrum$, to distinguish it from the 
$partial$ $spectra$ of individual ions. Recently, in the
last 10 years, the two major characteristics of the $knee$ 
have lost some of their fundamental importance since measurements of the energy
spectra of single ions, in large energy bands, have been reported [23, 24, 25] and 
their $knees$
differ from that of the complete spectrum. Thus, nowadays 
 the particular energies at which the changes
 of the spectral indexes of Proton and Helium take place
are known, and the energy spectra  as well. For the ion group CNO,
for Silicon and for Iron the $knees$ have not yet been observed [24, 26] though they
 are foreseen by many theorists.  Let us notice explicitly 
that the measured energy spectra of Helium, CNO elements and Iron between
 $10^{14}$-$ 10^{17}$  $eV$
are thoroughly
different from the complete spectrum. This spectrum has a constant slope 
with $\gamma$ = $3$ in the interval $ 3 \times 10^{15}$ - $ 5 \times 10^{18}$ 
$eV$ while the helium spectrum exhibits a slope depending on the 
energy. These facts
 allow to state, for example, that cosmic-ray intensity beyond the knee, for three
energy decades, is not due 
to the contributions of the intensities of many ions, each ion with a parallel 
slope $\gamma$ = 3.

\par As a long series of measurements indicate,  but particularly those from
Jacee [27, 28] and Runjob [29], the spectral index of Proton and Helium is $2.72$, fairly
constant in the range $10^{10}$ - $ 10^{15}$ $eV$. Taking this datum 
as a valuable reference and extrapolating to
 higher energies beyond the knee, it immediatly 
results a gap between the computed and 
 measured intensity of the cosmic radiation.
\begin{figure}
\centerline {\epsfig {figure=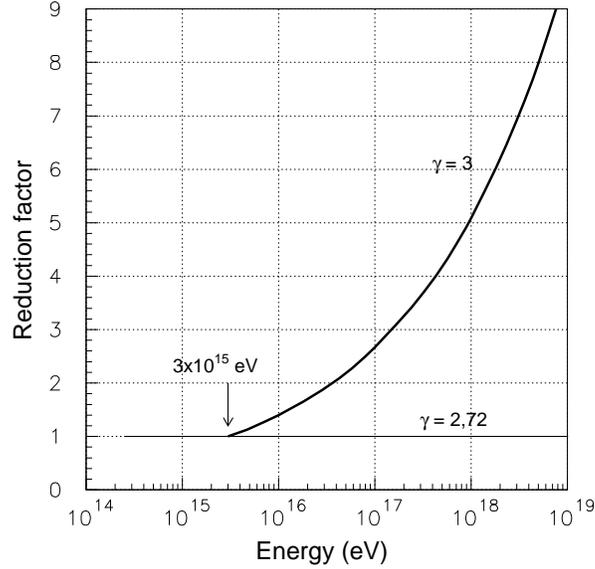,width=8cm}}
\caption{\it Deviation of the cosmic-ray intensity from an ideal energy spectrum
with an index $\gamma$ = $3$ in the energy range $10^{15}$ - $10^{19}$  $eV$.
The deviation is expressed as a reduction factor in the vertical axis. The initial
point, with no deviation, is placed at $ 3 \times 10^{15}$ $eV$ 
the nominal $knee$ energy of the $complete$ $spectrum$.
\label{fig:fig20}}
\end{figure}
Fig.$\,$\ref{fig:fig20} reports this gap, 
expressed as a reduction factor $versus$ energy.
 One should note that in the three energy decades $10^{15}$ - $ 10^{18}$ $eV$ 
this reduction factor is quite modest, not reaching the value 
of 8 at $ 10^{18}$ $eV$.
 The smallness of this factor should be 
compared with the 
enormous reduction factor, about $10^9$, of the complete spectrum in the same 
energy range. The comparison of these two different reduction factors 
 suggests interesting deductions developed in a recent study [7] and the 
irrefutable implication (see for example ref.[30])
that the knee is a small perturbation of the $complete$ $spectrum$.

\par It is customary debating the experimental data around the $knee$,
to multiply the complete spectrum by a factor  $E^{\gamma}$ 
so that changes in the spectral index, and eventually,
peaks or dips are magnified.
\begin{figure}
\centerline {\epsfig {figure=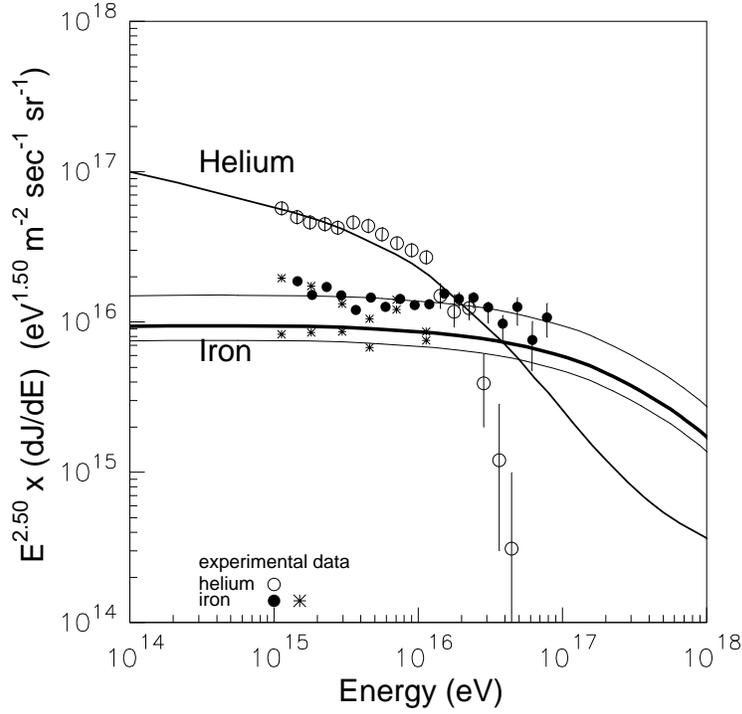,width=10cm}}
\caption{ Comparison of helium and iron spectra measured by the Kaskade and Eas-top
experiments with the results of this calculation.
Helium (void circles)
and iron (full circles) spectra from Kaskade ([24], [25]) and iron (stars) spectrum from 
Eas-top [26] experiments.
 Measurements of the Helium spectra from Kaskade [25] 
resulting in lower intensities are not shown. 
Their extrapolation at $10^{18}$ $eV$ would result in a major conflict
with the experimental data  
of the Haverah Park and Hires experiments.
Measurements of the Iron spectra
from Kaskade [24] resulting in lower intensities (from a factor $3$ to $4$) than those
shown here, incompatible with the maximum excursions of the Eas-top 
data [25], are not shown.
The thin (helium) and thick (iron) curves are the computed
spectra uncorrected for the galactic wind.
The thin curves below and above
the thick line (computed iron spectrum) suggest the uncertainty in the normalization
adopted for the comparison. The normalization of the calculation
is explained in the text. 
\label{fig:fig21}}
\end{figure}
\begin{figure}
\centerline {\epsfig {figure=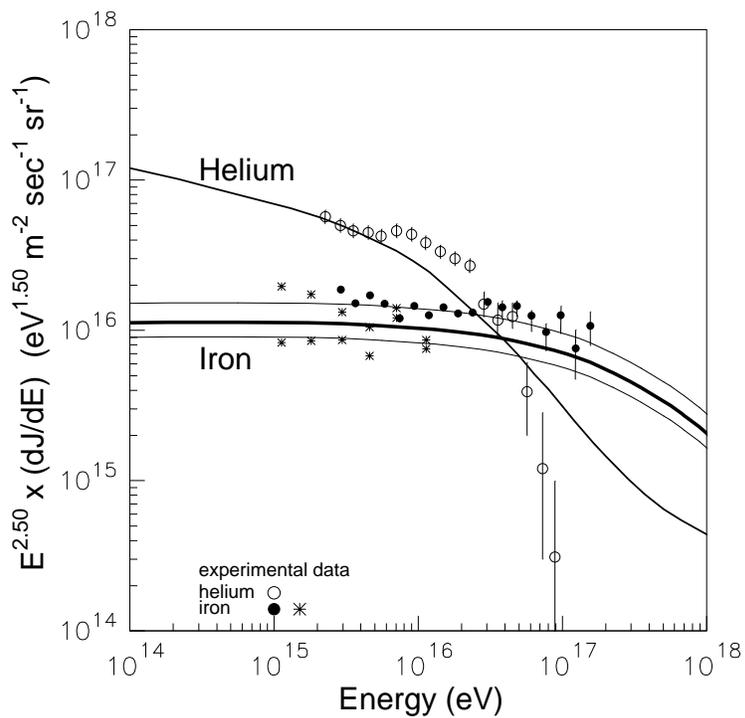,width=10cm}}
\caption{ Comparison of helium and iron spectra measured by the Kaskade and Eas-top
experiments with the results of this calculation. In order to assess the stability 
of the results
the energy of the experimental data has been arbitrarily shifted by a factor 2. 
The agreement is slightly improved. The computed spectra 
are the same as in fig.$\,$\ref{fig:fig21}. 
\label{fig:fig22}}
\end{figure}
\begin{figure}
\centerline {\epsfig {figure=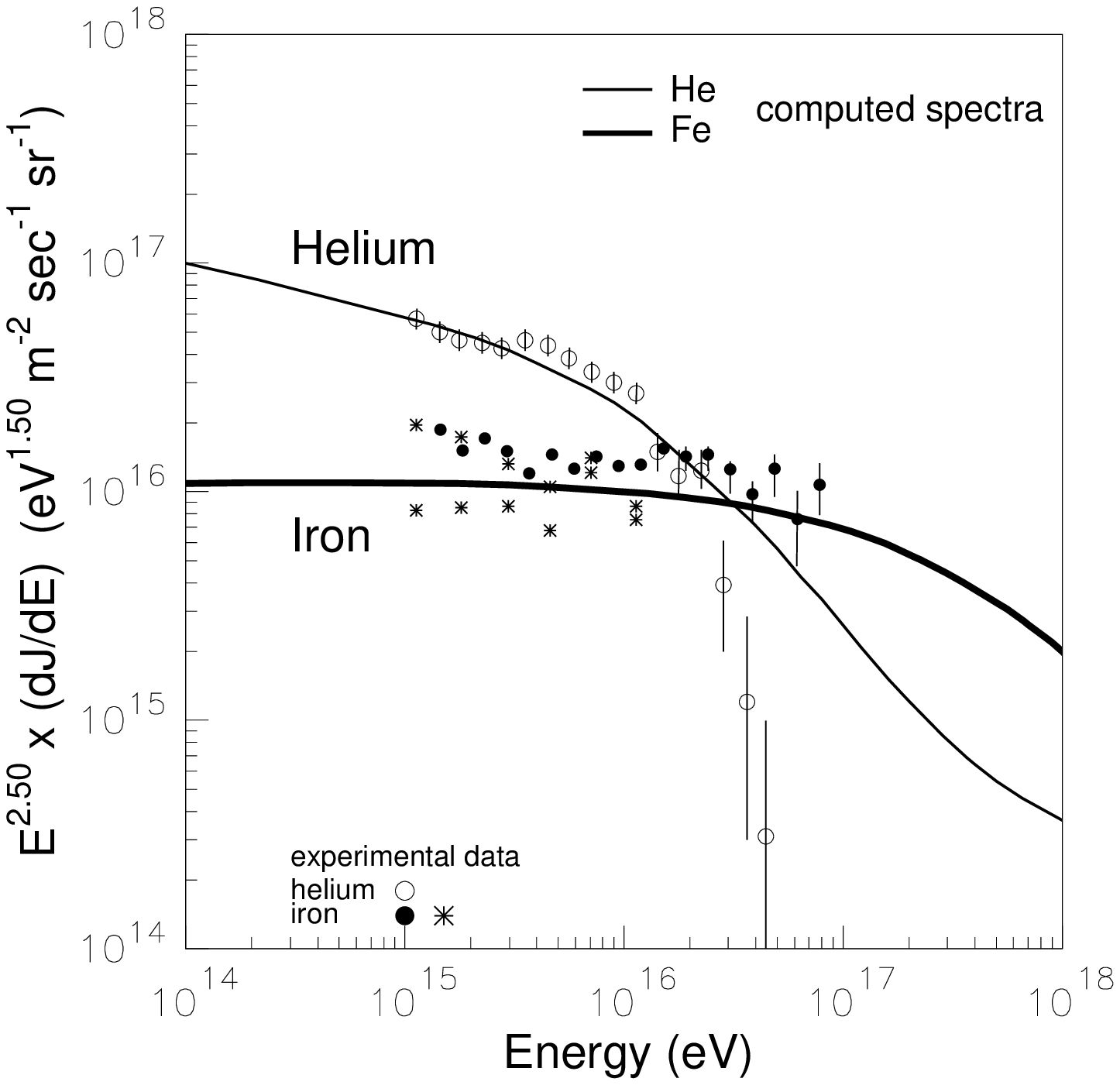,width=10cm}}
\caption{ Comparison of helium and iron spectra measured by the Kaskade and Eas-top
experiments with the results of this calculation. The effect of
the galactic wind is included in the two computed spectra which better conform
to the data.
\label{fig:fig23}}
\end{figure}
In fig.$\,$\ref{fig:fig21}  the measured Helium spectrum [24] is reported along
with the computed one (solid thick line). 
The computed curve is the same displayed in fig.$\,$\ref{fig:fig15} in a narrower
energy interval, to better focalize on the experimental data. \par The measured helium 
intensity is normalized to the computed one at the energy of  
$ 10^{15}$ $eV$. This normalization
 aknowledges and takes advantage of the 
$plateau$ of $n_g$ versus energy below $ 10^{14}$ $eV$ and of the measured
 spectral index of 2.7
for Helium. 
In the same fig.$\,$\ref{fig:fig21} is reported the computed 
iron spectrum (very thick line ) and 
 the iron data multiplied by $E^{\gamma}$
with $\gamma$ = $2.5$, being the value of the spectral index measured
below $ 10^{15}$ $eV$ [31-34].
 These data are limited to the maximum energy of
 $ 8 \times 10^{16}$ $eV$  and up to this energy, as already remarked, 
 no iron $knee$ is observed.
The iron spectrum measured by the Eas-top experiment at the Laboratori Nazionali del
Gran Sasso  [35], exhibits
the same trend.  Other important experiments [36,37] measuring helium spectra do not have 
sufficient energy 
band for a pertinent comparison here. 
\par The computed iron 
spectrum in fig.$\,$\ref{fig:fig21}  has been reduced
by a factor $2$ with respect to that of Helium. This factor takes into 
account  the different  powers of helium and iron sources 
in the Galaxy. In fact, by a simple comparison  
of the results of many experiments [31-37], in three energy decades, it emerges that
the Helium-to-Iron flux ratio is rather stable, amounting to about $2$. It ranges
from 2.8 at
 $ 5 \times 10^{11}$ $eV$ decreasing to $1.9$
around $10^{15}$ $eV$.
This value has been derived with no particular screening of the experimental data nor 
with a sophisticated analysis; it  is however
adequate for the data normalization here. What is silently implied by this normalization 
 is that some physical
processes (particle injection at the sources, acceleration mechanisms, reacceleration, 
propagation effects
others than those considered here, etc. ) 
regulating the intensity and energy spectra of cosmic rays are remarkably stable with 
energy in the range  $ 10^{11}$-$ 10^{15}$ $eV$.  
Renouncing to this normalization, a major
 number of difficulties would 
arise since different mechanisms, probably depending on energy, should match
together in such a way to generate an approximate constant He/Fe flux ratio, which is
unlikely.

\par The physical quantities in the vertical and
horizontal axes of fig.$\,$\ref{fig:fig21} have been arbitrarily altered
to test how critical is the accord; thus,
the energy of the experimental data [24, 34], is multiplied 
by a factor $2$ and the Iron intensities shifted upward  by the $15\%$ and $20\%$. 
 Fig.$\,$\ref{fig:fig22}  reports the results of this computational game and
testifies the great and amazing stability of the accord
without the effect of the galactic wind.
 Fig.$\,$\ref{fig:fig23}  reports the computed spectra with a galactic
wind of velocity 12 $km$/$s$. This is the same wind velocity adopted at very low energy,
which provides a better agreement between computed and measured B/C flux 
ratio (see for example ref.[18]).

\par The results in fig.$\,$\ref{fig:fig21}  indicate
 that the helium $knee$ is due to the fall of intensity caused by the 
particular value of the magnetic field strength in the Galaxy, limited to
  1-7 $\mu$$G$. Changing the value of the field strength, a shift in the position
of the $knee$ along the energy axis
is calculated, as vividly show the curves in fig.$\,$\ref{fig:fig10}. 
 Also important is the rise of nuclear cross sections as it stands out by comparing 
the results shown in fig.$\,$\ref{fig:fig13} 
 and fig.$\,$\ref{fig:fig15}.
Of capital importance 
in this conclusion, is the assumption that cosmic rays are galactic at the energies
considered. 
The computed helium knee agrees with the experimental data up to the energy 
of $ 5 \times 10^{16}$ $eV$. The helium data (three data points) above this energy 
 seem to fall too steeply.
The relevance of this deviation has to be compared with the large 
error bars and severe uncertainties
in the experimental data [25] beyond the energy of $ 8 \times 10^{16}$ $eV$.
In addition, the extrapolation of these data [25]  at $10^{18}$ $eV$ leads to an
 irriducible conflict with the outcomes 
of the Haverah Park and Hires experiments around $10^{18}$ $eV$, as discussed below.

\par Let us compare this explanation
 of the $knee$ with some previous attempts;
 these may be subdivided 
in six classes, depending on the physical mechanism invoked. ($I$) particle propagation
inside the disk does not allow a uniform containment, as the energy increases,
 because the
magnetic field strength is finite, amounting to about $1$-$7$ $\mu$$G$ [38 - 43].  
\quad ($II$) The acceleration mechanisms
of the galactic sources[44 - 49] become inefficient beyond a maximum energy, denoted here
$E_{max}$, related to the specific acceleration process called into play.   
Beyond $E_{max}$ a decline of the intensity at Earth should be observed. 
The regions
where cosmic ion suffer acceleration are supernovae remnants, pulsar 
atmosphere or, possibly, the entire Galaxy through reacceleration in the 
galactic wind [45].  ($III$) Cosmic-ion interactions with nuclei in the Earth atmosphere, 
in particular reactions channels, would yield exotic secondary
particles that might escape an efficient detection[50, 51] generating the $knee$. 
 ($IV$) Sites and mechanisms which produce $\gamma$ bursts in the galactic
 halos [52, 53]. 
($V$) Cosmic-ion interactions with photons present around the sources [54], ($VI$) 
or with background neutrinos in particular sites
 [55, 56] would induce effects in the intensity at Earth to account for the knee.

\par Most attempts become obsolete once the measurements of
individual knees are acknowledged, since the positions of the bends along the 
energy axis differ from that of the complete spectrum at about  $ 3 \times 10^{15}$ 
$eV$, where some tentative explanations were addressed or tuned. 
Moreover, the explanations ($III$), ($V$), ($VI$) require unobserved
particles or astrophysical conditions whose physical reality
is, at variance,  not proved or uncertain or very uncertain; as a consequence, 
calling them  to account
for the $knee$ problem seems hazardous. \par The solution of the knee problem
discussed here belongs to the class $I$. It is based on well established observational
facts incorporated in a limited set of parameters. This solution exhibits, 
step by step, how these parameters  forge 
the bend and the fall of the  Helium and Iron intensities
 between $ 10^{15}$-$10^{18}$.
\par Since the computed Helium and Iron $knees$ fairly join the available experimental data,
as shown  in fig.$\,$\ref{fig:fig23}, 
a small margin is left for other physical phenomena
to account for the origin of the $knees$. In particular, from this investigation 
follows that acceleration mechanisms,
ion filtering at injection to the galactic sources or other phenomena, are of little
 importance
for the $knee$ explanation. This conclusion reinforce a similar one [7], 
based on observational and theoretical arguments, indicating that acceleration mechanisms
have a negligible effect in generating the $knees$, regardless 
of the site and specific acceleration process invoked.
\par Finally, we cannot fail to mention that the results of this study demonstrate
the existence of a robust  extragalactic component in the energy decade 
$10^{17}$ - $ 10^{18}$ $eV$.

\par If the
 energy spectra of Helium and Proton
will continue to decrease as suggested by the experimental data [22,24], 
(The Tibet-b experiment [37]  
reports $\gamma$ = 3.06 for protons and a bend close to $ 10^{14}$ $eV$),
the proton fraction will become 
insignificant at  $ 10^{18}$ $eV$.
 Measurements of the proton fraction around
 $ 10^{18}$ $eV$ 
of Haverah Park and Hires experiments [57, 58, 59] indicate a value
of about 40 per cent, much higher than  the fractions extrapolated from the 
 data quoted above [24]. Therefore, the proton and helium intensities 
above  the $knee$ energies are expected to increase again. The particular 
energies at which the 
ascents would take place and their steepnesses, for each ion, are still experimentally
vague. According to this study it is 
impossible, using only galactic sources,
to generate  large fractions of Protons, Helium, CNO or Iron around the energy 
of $ 10^{18}$ $eV$ 
 because the computed intensities 
in fig.$\,$\ref{fig:fig15} invariantly fall with energy 
with a spectral index greater than $3$. 
 An obvious consequence, 
given the existence of the complete spectrum
of cosmic rays which falls with a constant spectral index of $3$ in this energy range, 
 is that additional sources of Helium and Proton efficiently penetrate the 
local galactic zone at energies slightly above the $knees$. These additional sources
have to compensate the fall of the $knees$ in order to recover the index $\gamma$ = $3$
of the complete spectrum. But, if
these additional sources are not placed in the
Galaxy,  they must be extragalactic. 
\vskip 0.9cm

{\bf Acknowledgments}
\vskip 0.2cm
\par We are indebted to Prof. G. Navarra from Eas-top Collaboration
for providing us with the exact values of the published experimental data 
on Iron displayed in 
figure 21.
 
\section{Appendix A \quad A reference frame for spirals}

It is preferable, though unnecessary, to express some
features of the galactic basins, in a  reference frame that explicitly  recognizes
the dominant role of the regular component of the galactic magnetic field.
Fig.$\,$\ref{fig:fig24} illustrates this frame that has been used   
in fig.$\,$\ref{fig:fig7} and fig.$\,$\ref{fig:fig11}. 
\par The length of the spiral arc between the Bulge and the disk periphery 
( $r$ = $15$ $kpc$) is given by:
$$ {(R_d - R_b) \over cos(\theta_L)} = {s} \quad \quad \quad (4)$$
where $R_d$ is the disk radius, $R_b$ that of the Bulge, $\theta_L$
the opening angle defined in fig.$\,$\ref{fig:fig24} and $s$  the length of the logarithmic 
spiral  between the radii $R_b$ and $R_d$. 
The total angular opening of the spiral viewed from the galactic center
is chosen to be  $3$/$4$ of a complete turn i.e. $\times 2\pi$. With this choise
$s$ = $40.73$ $kpc$. The factor $3/4$ is the number of turns of the logarithmic spiral 
about the galactic center which, in this case, is less than 1.
The angle $\theta_L$ beween any radius drawn from
the galactic center and the logarithmic spiral crossing the local zone 
is given by:
$$ tg (\theta_L) = {3 \over 4} \times {2\pi} \times {1 \over ln(R_d / R_b)} \quad
 \quad \quad (5) $$
resulting in $\theta_L$ = $74.332$ $degrees$.
Taking a typical lateral extension of the basin of $300$ $pc$,
a length-to-width basin ratio $L_B$/$W_B$ of about $10^2$ is obtained. In this work the
 high value of the $L_B$/$W_B$  ratio has practical consequences in the 
description of the basin  for  graphical 
representation and for computing time  as well.
\par Let us define a refence frame, $R_N$, in the galactic midplane
with origin $N$ and axes $x_L$ and $y_L$ ( see figure 24). In this frame any source
 position placed
at the arbitrary point $P$ has coordinates $X_P$ and $Y_P$.  The $X_P$ coordinate is defined 
as the normal drawn from the point $P$ to the principal field line, the 
segment $PN$ of length $\Delta{R_\perp}$ . The $Y_P$ coordinate is zero by definition.
 Let us call  $R_Z$ a new frame
with origin $Z$ in the local galactic zone with x and y axes having
the same orientations as the axes $x_L$ and  $y_L$ in the frame $R_N$.  
 The $y$ coordinate of the point $P$ is the length of the spiral arc $ZN$
counted from the $local$ $zone$ and the corresponding $x$ coordinate is always 
equal to $X_P$. 
In this coordinate system the disk edge along the principal field line
 has a $y$-coordinate of +$24.07$ $kpc$ while the bulge
edge -$16.66$ $kpc$. Fig.$\,$\ref{fig:fig24} defines and illustrates some 
interesting variables in the frames $R_N$ and $R_Z$. Thus, the $x$ and $y$ coordinates of any point $P$,
in the frame $R_Z$ are:
\begin{figure}
\centerline {\epsfig {figure=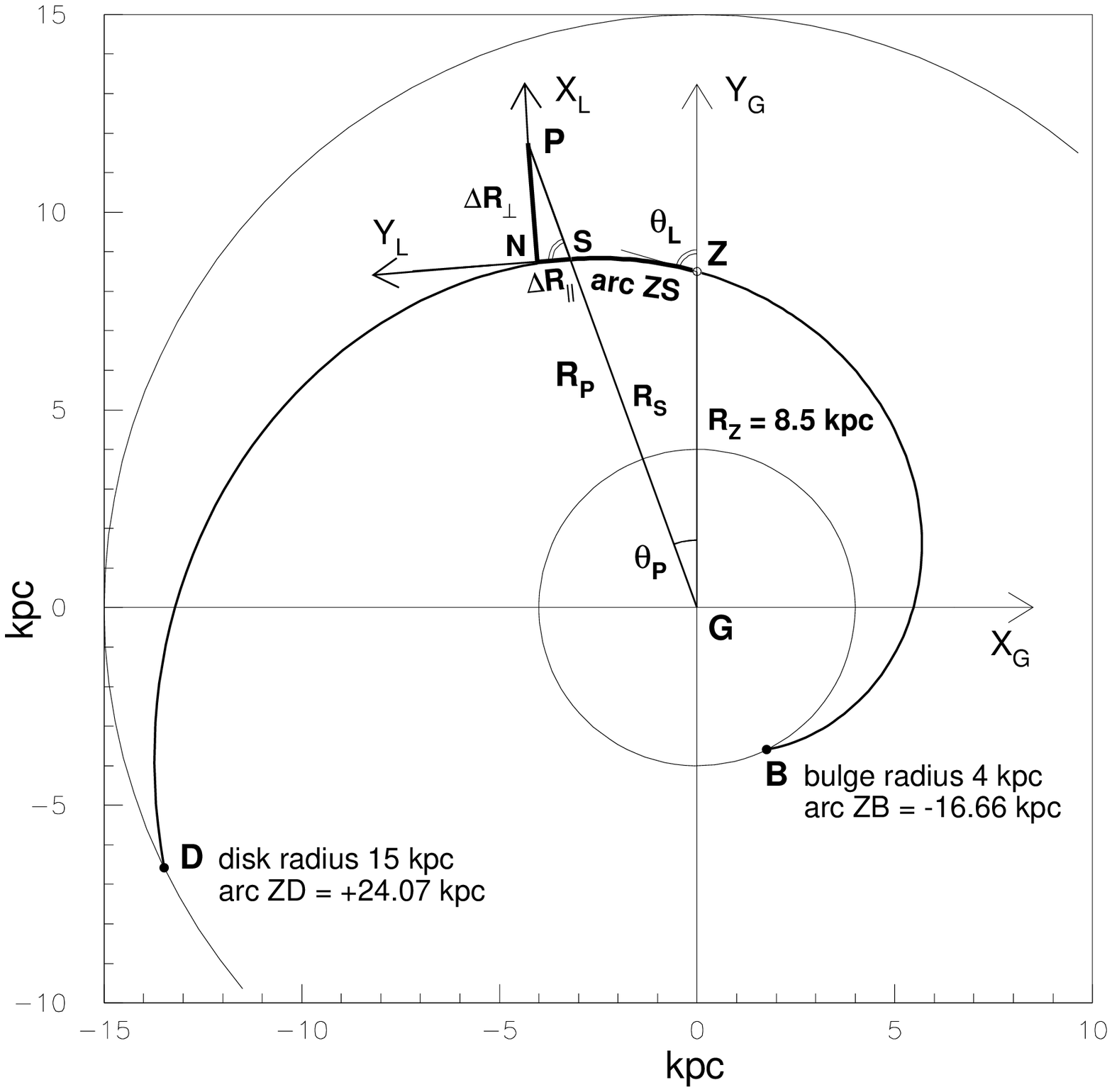,width=10cm}}
\caption{ Definition of the reference frame bound to the particular spiral
rooted in the bulge frontier and terminating in the disk radius
with a length $s$ of $40.73$ $pc$.
The point $Z$ denotes the position of the local zone while the point
$P$ that of a generic source. The angle $\theta_L$ between the spiral and
any segment departing from $G$ intercepting the spiral is constant.
The thick line represents the principal field line
which is an arc of a logarithmic spiral.
 The contour plots
shown in fig.$\,$\ref{fig:fig7} are given in the frame $R_Z$ defined in the text.
\label{fig:fig24}}
\end{figure}
$$ x = X_P = \Delta{R_\perp} \quad \quad \quad (6) $$
$$ y = arc (ZS) + \Delta{R_\parallel} \quad \quad \quad (7) $$

The quantity $\Delta{R_\parallel}$ (see fig.$\,$\ref{fig:fig24} ) is the $Y_L$ coordinate of the
point  $S$ which is the interception between the spiral and the segment $GP$.
\par The spatial distribution of the sources in the frame $R_Z$ 
have an approximate symmetry in +-$y$ and in +-$x$ as displayed in figure 6 and 7. 
Small distortions
of these symmetries are expected, for example, on the fact that 
the magnetic field strength shown in Fig.$\,$\ref{fig:fig3}  is not completely symmetric along
the principal field line.

\section{Appendix B \quad Handling ion trajectories in the Galaxy}

Ion type, its energy and nuclear cross sections are always specified at the
beginning of each trajectory  simulation. Particle
 propagation through the galactic volume initiates 
at some point, of coordinates $r$, $z$, $\phi$, and terminates at a final point specified 
by the events described in Section 4. The initial point
is the source position, or eventually, the nuclear collision point, where 
the secondary particles are generated. Nuclear decays, presently dormant, are also 
handled by the code $Corsa$. 
The trajectory links the source to the final point.

\par Trajectories are subdivided in short and long segments. The long segments are
those contained in the regular field. They are  sampled according to an exponential
 function
with a mean value of $125$ $pc$ [12]. This quantity is regarded as $coherence$
$length$ of the field. The short segments are those developed inside
the magnetic cloudlets.
 The full trajectory is a sequence of long
and short segments. Several thousands of segments are involved in a typical
trajectory in the disk volume, and hundreds of thousands in the Halo.
Each segment is regarded as a helix of constant radius. 

\par In a trajectory segment the magnetic field strength is taken constant and equal
to its value, at the beginning of a segment, 
in spite of the variable field strength, as displayed 
in Fig.$\,$\ref{fig:fig3} .
Since a typical trajectory is dispersed in a large volume, this simplification,
on average, do not affect the quantities calculated in the present paper.
The helices shown in figure 4 are drawn under this hypothesis.

\par A further simplification adopted at low energy, below $ 10^{14}$ $eV$/$u$, 
in order to reduce 
the computing time, is to identify
the helix axes with the trajectory axes. The ion gyroradius
is quite small at low energy compared to the size of the disk. Hence, the 
resulting helices have many turns.
In this case the direction of motion of the ion at the final point of the
trajectory segment ( AB in Fig.$\,$\ref{fig:fig4} )
is uniformly distributed
about the helix axis, as shown by the cone in Fig.$\,$\ref{fig:fig4} .
 Thus magnetic
 cloudlets, as they are randomly oriented, not only scatter particles 
in all directions, but they can invert the direction of the 
motion. 
\par At very high energy,  above $ 10^{14}$ $eV$/$u$, depending on the 
number of turns of the ion  in the helix (less than $1$, 
a few or many turns),  different 
algorithms are at work for shaping the physical trajectories.
Fig.$\,$\ref{fig:fig4} shows helices of three 
different radii to qualitatively illustrate some categories of trajectories. 
The segment $AB$ in Fig.$\,$\ref{fig:fig4}  
represents  either the size of a cloudlet or that of a regular field segment. 
\par As an example, the particle escaping with a
velocity vector $\alpha$ tangent to the helix arc in $C$
illustrates how the simulation algorithms handle very energetic particles 
leaving a field region cell.
Note that
the length of the segment $AC$ equals the size of the magnetic field 
region. 
\par At some energy, particles travel less than one turn, along the
 helicoidal path, in a magnetic 
field region of specified strength. 
Using an average field strength of  $1$ $\mu$G  and an arbitrary energy of
$ 10^{16}$ $eV$, 
the helium gyroradius is $11.1$ $pc$.
 
\par It is of importance to determine the energies  where the invertion 
of motion operated by  magnetic field 
 ceases completely signaling an almost rectilinear propagation of cosmic ions.  
This takes place for $1$ $\mu$G field strength at
$ 10^{16}$ $eV$  for Helium and $ 10^{18}$ $eV$ for Iron. Beyond these energies, 
helium trajectories are fractions of helix arcs, becoming almost straight line
segments at higher and higher energies. In this case the simulation code
generates the exact forms of the trajectories, as far as the
 magnetic field strength
and its size, in the sequence of long and short trajectory segments,
corresponds to the correct representation of the physical reality.

Fig.$\,$\ref{fig:fig4}  also shows that the axes of two
contiguous trajectory segments  for large and differing gyroradii are 
significantly shifted.



\begin{thebibliography}{99}

\bibitem{1}  Codino A. and Plouin F., 
             The Astrophysical Journal, 639, 173, 2006.
\bibitem{2}  Codino A. and Plouin F., 
             in Proc. 28th Int. Cosmic Ray Conference,
             Tsukuba, Japan, Session OG.1, p. 1977 (2003).
\bibitem{3}  Codino A. and F. Plouin,   
             A unique mechanism generating the knee and the ankle
             in the local galactic zone, in preparation.
\bibitem{4}  Brunetti M. T. and Codino A., 
             The astrophysical Journal, 528, 789, (2000).
\bibitem{5}  Codino A., Conference proceeding, 
             Vulcano Workshop, 439 (1998).
\bibitem{6}  Codino A., Brunetti M. T. and Menichelli M., 
             Proc. 24th ICRC, Rome, Italy, 3, 100 (1995).
\bibitem{7}  Codino A., 
             Proc. 29th ICRC, Pune, India, (2005).
\bibitem{8}  Codino A., 
             Proc. 29th ICRC, Pune, India, (2005). 
\bibitem{9}  Gaisser T. K., 
             Cosmic Rays and Particle Physics,
             Cambridge, Cambridge University Press, 1990.
\bibitem{10} Bell M. C. et al.,
             J. Phys. G., 7 (1974).
\bibitem{11} Vallee J. P., 
             Fundam. Cosmic physics, 19, 319 (1998).
\bibitem{12} Chi X. and Wolfendale A. W., 
             J. Phys. G, 16, 1409 (1990); 
             Osborne J. L. et al., 
             J. Phys. G,  6, 421 (1973).
\bibitem{13} Giacalone J. and Jokipii J. R., 
             The Astrophysical Journal, 520, 204 (1999).
\bibitem{14} Stecker S. W. and Jones F.C., 
             Proc. 12th ESLAB Symposium, 171 (1997).
\bibitem{15} Pohl M., Esposito J. A. {\it et al}, 
             The Astrophysical Journal, 507, p. 327 (1998).
\bibitem{16} Lerche I. and Schlickeiser R. ,
             The Astrophysical Journal, 239, 1089 (1980).
\bibitem{17} Zirakashvili V. N. et al., 
             Astronomy and Astrophysics, 311, 113-126 (1996).
\bibitem{18} Strong A. W. and Moskalenko I. V., 
             The Astrophysical Journal, 509, 212 (1998).
\bibitem{19} Amos N. A., Avila C., Baker W. F., Bertani M., Block M. M. 
             and Dimitroyannis D. A.,
             Phys. Rev. Lett. 63, 2784 (1989); 
             Abe F. et al., 
             Phys. Rev. D 50, 5550 (1994);
             Avila C. et al., 
             Phys. Lett. B 445, 419 (1999).
\bibitem{20} Codino A. and Vocca H., 
             Proc. 27th ICRC Hamburg, Germany (2001).
\bibitem{21} Clay R. W.  et al., 
             Proc. 25th ICRC Durban, South Africa, 4, 185 (1997).
\bibitem{22} Kulikov G. V. et al.,
             JEPT, 35, 635 (1958).
\bibitem{23} Amenomori M. et al., 
             Phys. Rev. D 62, 112002 (2000). 
\bibitem{24} Roth M. et al., 
             Proc. 29th ICRC, Tsukuba, Japan (2003).
\bibitem{25} Antoni  T. et al., 
             Astroparticle Physics, 24, 1 (2005). 
\bibitem{26} Navarra et. al., 
             Proc. 28th ICRC, Tsukuba, 1, 147, Japan (2003).
\bibitem{27} Asakimori K. et al., 
             The Astrophysical Journal, 502, 278 (1998)
\bibitem{28} Parnell T. A. et al., 
             Adv. Space Research, Vol. 9, No 12 (1989).
\bibitem{29} Derbina V. A. et al., 
             The Astrophysical Journal, 628, L41 (2005).
\bibitem{30} Stenkin Y. V., 
             Modern Physics Letters A, 18, No. 18, 1225-1234 (2003).
\bibitem{31} Mueller H. D. et al., 
             Adv. Space Research, Vol. 9, No 12 (1989).
\bibitem{32} Simon M. et al., 
             The Astrophysical Journal, 239, 712 (1980).
\bibitem{33} Juliusson et al., 
             The Astrophysical Journal, 191, 331 (1974).
\bibitem{34} Mueller H. D. et al., 
             The Astrophysical Journal, 374, 356 (1991).
\bibitem{35} Aglietta M. et al., 
             Astroparticle  Physics, 21, 584-596, (2004).
\bibitem{36} Huang J. et al., 
             Astroparticle Physics, 18, 637 (2003).
\bibitem{37} Amenomori M. et al., 
             Phys. Rev.  D 62, 112002 (2000). 
\bibitem{38} Ptuskin S. V. et al., 
             Astronomy and Astrophysics, 268, 726 (1993).
\bibitem{39} Kalmikov N. N. and Pavlov A. I.,
             Proc. 26th ICRC Salt Lake City, Utah, 4, 263 (1999).
\bibitem{40} Wdowczyk J. and Wolfendale A. W., Jour. Physics 
             G., 10, 1453 (1984).
\bibitem{41} Ogio S. and Kakimoto F., 
             Proc. 28th ICRC, Tsukuba 1,315 (2003). 
\bibitem{42} Swordy S. P., 
             Proc. 24th ICRC, Rome, 2, 697 (1995).
\bibitem{43} Lagutin A. A. et al., 
             Nucl. Phys. B (Proc. Supp.) 97,267 (2001).
\bibitem{44} Berezhko E. G. et al.,
             JETP 82, 1 (1996).
\bibitem{45} Jokipii J. R. and Morfill G. E., 
             The Astrophysical Journal, 312, 170, (1986).
\bibitem{46} Stanev T. et al., 
             Astronomy and Astrophysics, 274, 902 (1993).
\bibitem{47} Kobayakawa K. et al., 
             Phys. Rev. D 66, 083004 (2002).
\bibitem{48} Sveshnikova L.G. et al., 
             Astronomy and Astrophysics, 409, 799 (2003).
\bibitem{49} Erlykin A. D. and Wolfendale A. W.,
             J. Phys. G: Nucl. Part. Phys. 27,1005 (2001).
\bibitem{50} Nikolsky S. I., 
             Phys. Atomic Nuclei 63, 1799 (2000).
\bibitem{51} Petrukhin A. A., 
             Phys. Atomic Nuclei 66, 517 (2003).
\bibitem{52} Plaga R., 
             New Astronomy, 7, 317 (2002).
\bibitem{53} Wick S. D. et al., 
             Astroparticle  Physics, 21, 125 (2004).
\bibitem{54} Hillas A. M., 
             in 16th ICRC, kyoto, Japan, Vol.8, p.7 (1979).
\bibitem{55} Karakula S. and Tkaczyk W., 
             Astroparticle Physics, 1, 229 (1993).
\bibitem{56} Wigmans R. , 
             Astroparticle Physics,  19, 379 (2003).
\bibitem{57} Barrett M. L. et al., 
             in Proc. 15th ICRC, Plovdiv, Bulgaria, Vol. 8, p 172 (1977).
\bibitem{58} T. R. Gaisser et al., 
             Comments on Astrophysics, Vol. 17, pp 103-117 (1993).
\bibitem{59} Walker R. and Watson A. A.,
             J. Phys. G, 8, 1131 (1982).

\end{thebibliography}
\end{document}